\documentclass[aps,prd,twocolumn,showpacs,superscriptaddress,showkeys,bm,amsmath,amssymb,amsart,longbibliography,raggedbottom]{revtex4-1}

\usepackage{xcolor}
\usepackage{graphicx}
\usepackage{multirow}
\usepackage{textcomp}
\usepackage{soul}

\begin{document}
\title{Axion Haloscope Using an 18\,T High Temperature Superconducting Magnet}
\newcommand{\KAIST}{\affiliation{Department of Physics, Korea Advanced Institute of Science and Technology, Daejeon 34141, Korea}}
\newcommand{\SNU}{\affiliation{Department of Physics and Astronomy, Seoul National University, Seoul 08826, Korea}}
\newcommand{\IBS}{\affiliation{Center for Axion and Precision Physics Research, Institute for Basic Science, Daejeon, 34051, Korea}}

\author{Hojin Yoon}\KAIST
\author{Moohyun Ahn}\SNU
\author{Byeongsu Yang}\SNU\IBS
\author{Youngjae Lee}\KAIST
\author{DongLak Kim}\IBS
\author{Heejun Park}\IBS
\author{Byeonghun Min}\IBS
\author{Jonghee Yoo}\SNU\KAIST\IBS

\date{\today}
\begin{abstract}
{We report details on the axion dark matter search experiment that uses the new technologies of a High-Temperature Superconducting (HTS) magnet and a Josephson Parametric Converter (JPC). An 18\,T HTS solenoid magnet is developed for this experiment. The JPC is used as the first stage amplifier to achieve a near quantum-limited low-noise condition. A first dark-matter axion search was performed with the 18\,T axion haloscope\,\cite{Lee:2022mnc}. The scan frequency range is from 4.7789\,GHz to 4.8094\,GHz (30.5\,MHz range). No significant signal consistent with Galactic dark matter axion is observed. Our results set the best limit of the axion-photon-photon coupling ($g_{a\gamma\gamma}$) in the axion mass range of 19.764 to 19.890\,$\mu$eV. Using the Bayesian method, the upper bounds of $g_{a\gamma\gamma}$ are set at 0.98$\times|g_{a\gamma\gamma}^{\text{KSVZ}}|$ (1.11$\times|g_{a\gamma\gamma}^{\text{KSVZ}}|$) in the mass ranges of 19.764 to 19.771\,$\mu$eV (19.863 to 19.890\,$\mu$eV), and at 1.76 $\times|g_{a\gamma\gamma}^{\text{KSVZ}}|$ in the mass ranges of 19.772 to 19.863\,$\mu$eV with 90\% confidence level, respectively. We report design, construction, operation, and data analysis of the 18\,T axion haloscope experiment.}
\end{abstract}
\maketitle

\section{Introduction}
\par The existence of dark matter is well established by astrophysical observations. However, the properties of dark matter are barely known. The simplest assumption of dark matter is that it is non-luminous, non-absorbing, and rarely interacts with ordinary matter. Based on the analysis of the structure formation of the galaxy, the majority of dark matter should be non-relativistic. There is no lack of dark matter particle candidates, including supersymmetric particles, primordial black holes, and axions. In the case of thermal relics, the mass of dark matter particles must be above a keV to be involved in galaxy formation. However, this argument does not constrain the mass of non-thermal relic particles. Especially the axion is one of the most prominent non-thermal relic dark matter candidates\,\cite{PRESKILL1983127, ABBOTT1983133, DINE1983137, Aghanim:2018eyx}. \\

\par Axions were originally postulated to solve the strong-CP problem in quantum chromodynamics (QCD)\,\cite{kim2010axions, Peccei:1977hh, Wilczek:1977pj, Kim:1979if, Dine:1981rt}. A global Peccei-Quinn U(1) symmetry breaking mechanism was suggested to solve the problem. As a result of the symmetry breaking at the scale of $f_a$, axions are inevitably produced\,\cite{Peccei:1977hh,Weinberg:1977ma,Wilczek:1977pj}. Axions were massless in the early Universe when the temperature was above the QCD phase transition scale. The axion's mass was formed as Universe cooled below the QCD scale. The majority of earlier axion models, which assumed $f_a$ to be an electroweak symmetry breaking scale ($\Gamma_{\text w} \simeq$ 247\,GeV), were discarded by experimental constraints. However, axion models with $f_a \gg \Gamma_{\text w}$ have survived over the current experimental bound. \\

\par The axion coupling to ordinary Standard Model particle is predicted by theoretical models. Among them, two prototype models are prominent due to their symmetry-breaking properties. They are the Kim-Shifman-Vainshtein-Zakharov (KSVZ)\,\cite{Kim:1979if,Shifman:1979if} and Dine-Fischler-Srednicki-Zhitnitsky (DFSZ)\,\cite{Dine:1981rt,Zhitnitsky:1980tq} invisible axion models. In the KSVZ model, QCD axion coupling to fermions is negligible as it vanishes at tree level. The model invokes electrically neutral hypothetical heavy quarks to carry $U(1)_{\text{PQ}}$ charges; hence it is called the ``hadronic axion model." In the DFSZ model, QCD axion couples to fermions at tree level, and the model requires an extra Higgs doublet. The fermions carry $U(1)_{\text{PQ}}$ charges; hence it is called the ``fermionic axion model." The interaction strengths of these two models are well parameterized for experimental tests. In particular, an axion in the mass range of $1\sim100\,\mu$eV is an ideal dark matter candidate. Experimental tests for these dark matter axions are within reach using recently developed detector technologies.\\ 

\par The {\it axion haloscope} is an outstanding detector technology for the dark matter axion search\,\cite{sikivie1983experimental,sikivie1985detection}. In an axion haloscope, a cylindrical resonant radio-frequency (RF) cavity is deployed in the center of a strong axial B-field solenoid magnet. Dark matter axions in the local galactic halo may couple to the applied magnetic field and be converted to microwave photons. The microwave signal power in the cavity is 
\begin{eqnarray}
\label{eq:APower}
P^a = g^2_{a\gamma\gamma}\Big(\dfrac{\rho_a}{m^2_a}\Big) \omega_0 B^2_0 V C_{nlm}Q_0\frac{\beta}{(1+\beta)^2},
\label{eqn:pa}
\end{eqnarray}
\noindent where the axion-photon-photon coupling ($g_{a\gamma\gamma}$) and mass of axion ($m_a$) are unknown parameters. The kinetic energy profile of the local dark matter axions is assumed to be a Maxwell-Boltzmann (MB) distribution~\cite{PhysRevD.42.3572}, where the range of the velocities is from 110 to 360\,km/s, with the effective width of 250\,km/s. The mass density of axion dark matter ($\rho_a$=0.45\,GeV/cc) is from the dark matter halo model\,\cite{LEWIN199687, PhysRevD.42.3572}. In the haloscope, the axion-photon conversion power is proportional to the square of the external magnetic field strength ($B^2_0$), the effective volume of the cavity ($V$), the cavity form-factor ($C_{nlm}$), the unloaded quality-factor of the cavity ($Q_0$), and coupling ($\beta$) between the RF antenna and cavity modes. A typical axion signal power using conventional instruments is an order of $\sim10^{-24}$\,W. One of the challenging factors of a haloscope is the narrow-band character of the searching frequency, which is a trade-off of the high Q-factor of the cavity. Therefore, the target frequency must be re-tuned every time to probe a new axion mass. The scan speed of the axion mass is often regarded as the figure of merit when designing an axion haloscope. The scan speed is given by $dm_a/dt \propto B_0^4 V^2Q_L/T_S^2$, where $V$ is practically constrained by the choice of the magnet bore size. The loaded Q-factor $Q_L=Q_0/(1+\beta)$ depends linearly on the scan speed. Therefore, the external magnetic field ($B_0$) and system noise temperature ($T_S$) are practically the most significant design parameters of an axion haloscope.\\

\par Axion Dark Matter eXperiment (ADMX) is currently the only axion haloscope experiment that demonstrated the discovery potential of the invisible QCD axion dark matters for the DFSZ models with an unprecedented scan-speed\,\cite{Bartram:2020ysy,Braine:2019fqb,Du:2018uak,Asztalos:2009yp,Asztalos:2003px,Hagmann1998a}, while other experiments such as HAYSTAC\,\cite{Brubaker:2016ktl,Brubaker:2017rna}, QUAX-$a\gamma$\,\cite{Alesini2019a, Alesini2021a}, and CAPP experiments\,\cite{Lee:2020cfj,Jeong:2020cwz, Kwon2021a} are advancing their detector technologies to achieve competitive sensitivities. Recently, HAYSTAC collaboration reported a dark matter axion search result with a substantially improved amplifier noise performance using a squeezed-state receiver (SSR) technology. The results showed that the DFSZ axion search in the high mass region would be ultimately possible in the very near future\,\cite{Backes2021a}. \\

\par The axion haloscope presented in this report (CAPP18T) is built using two major technologies: (1) a High-Temperature Superconducting (HTS) magnet technology\,\cite{yoon201626t,jang2017design,Hahn2019Nature}, and (2) a Josephson Parametric Converter (JPC) amplifier which demonstrated a near quantum-limited low-noise performance\,\cite{bergeal2010phase, Bergeal2010JPC, LiuJPC2017} in the frequency range of interest. This paper reports on details of the first result of searching for invisible axion dark matter using the CAPP18T axion haloscope\,\cite{Lee:2022mnc}. This report includes design, construction, operation, and data analysis.\\

\begin{figure}[t!]
\centering
\includegraphics[width=.8\linewidth]{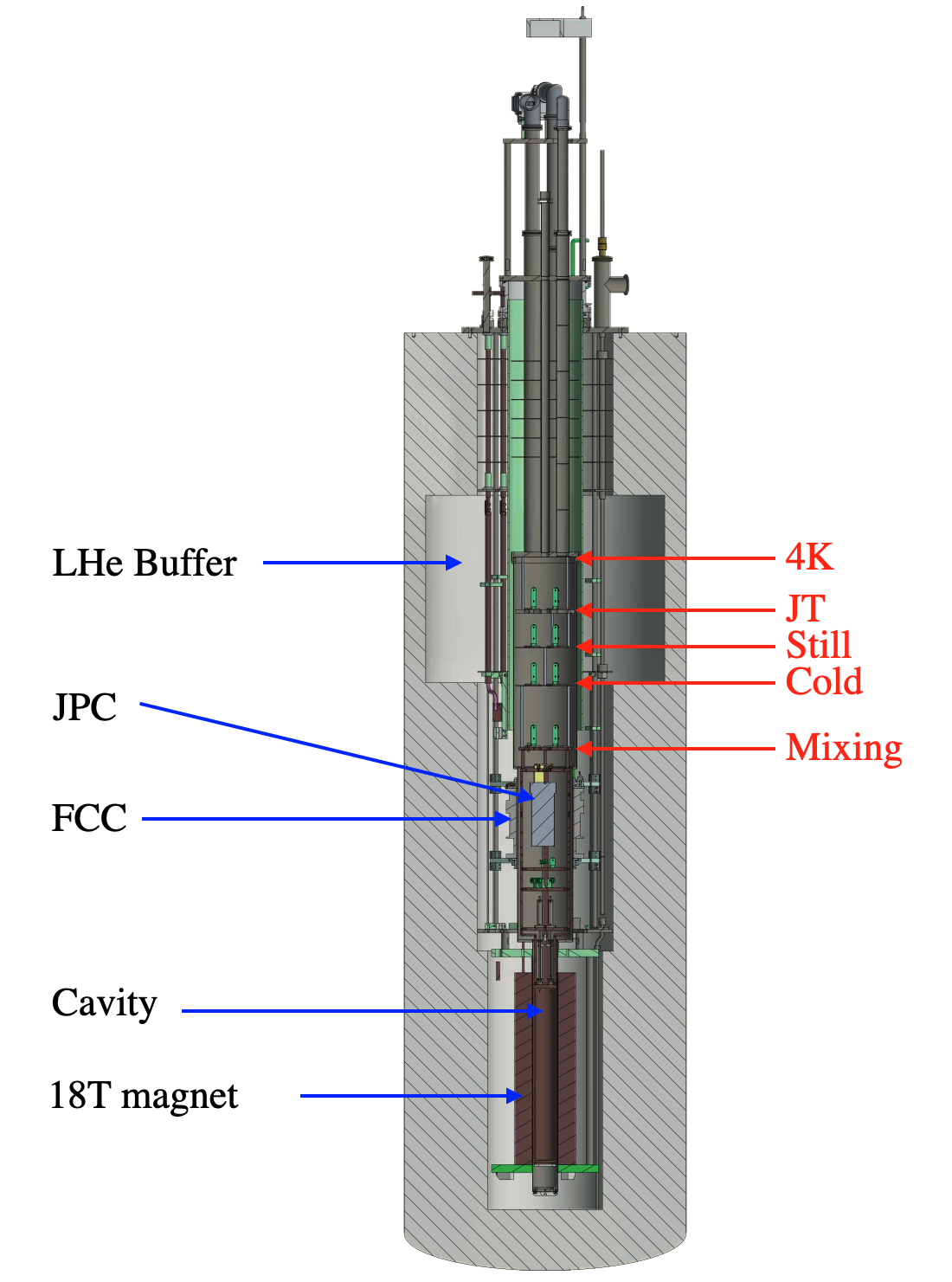}
\caption{Schematic view of the 18\,T axion haloscope. The 18\,T HTS magnet, and Field Cancellation Coil (FCC) are installed in a liquid Helium (LHe) bath. Other detector structures and elements are installed in an Inner Vacuum Chamber (IVC). The bottom part of the IVC, where an RF cavity is installed, is inserted in the bore of the 18\,T magnet. Dilution Refrigerator (DR) temperature stages are shown in red font as 4K flange (4K), Joule-Thomson stage (JT), Still, cold plate (Cold), and mixing chamber.}
\label{fig:18Texp}
\end{figure}

\section{18\,T Axion Haloscope}
\par The major constituents of an axion haloscope are a strong axial B-field solenoid magnet, a Field Cancellation Coil (FCC), an RF microwave resonant cavity, a Dilution Refrigerator (DR), and Low Noise Amplifiers (LNAs). FIG.~\ref{fig:18Texp} shows a schematic section view of the CAPP18T haloscope. A resonant cavity is installed in the center of the 18\,T HTS magnet. An FCC is installed to cancel the stray magnetic field from the 18\,T magnet. A JPC is located in the center of the FCC. RF signals from the JPC are further amplified by a second-stage cryogenic amplifier, a High Electron Mobility Transistor (HEMT). The cryogenic RF receiver chain is guided to the room temperature (RT) stage. RF signals are further amplified and processed in the RT-RF chain. A Data Acquisition (DAQ) system consists of an Analog-to-Digital-Converter (ADC) and a dead-time-free data processing software package. An online slow control monitor displays the essential detector parameters in real-time. This section describes the details of these detector components.\\

\subsection{18\,T High Temperature Superconducting Magnet}
\begin{figure}[t!]
\centering
\includegraphics[width=.5\linewidth]{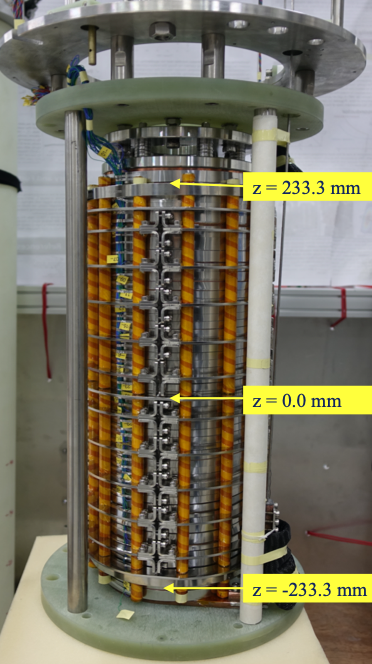}
\caption{A photograph of the 18\,T HTS magnet. }
\label{fig:18Tmagnet}
\end{figure}
\par A magnet is one of the most crucial components of an axion haloscope, as the scanning speed of the axion mass is proportional to the fourth power of the applied axial magnetic field. In general, Low-Temperature Superconductors (LTS) are used to build a high-field magnet. The highest magnetic field that an LTS has produced is 23.5\,T\,\cite{Bruker_ann}. Magnetic fields beyond 25\,T can hardly be achieved with the LTS technology, due to the limited critical field that LTS material can sustain~\cite{LTS_FIG}. The HTS magnet technology overcomes this limitation and potentially reaches beyond 50\,T. Recently, the HTS magnet technology has advanced in developing mechanically strong HTS tapes, which can persist under large hoop stress of high current density. Especially, REBCO (Rare Earth Barium Copper Oxide) superconductors have demonstrated high critical temperature T$_c$ (91\,K), high current density ($>$100\,A at 4.2\,K), high critical field ($\sim$100\,T), and large tension ($>$100\,MPa). These high-performance REBCO tapes are used to develop a strong magnetic field in a compact volume\,\cite{hahn2011hts, hahn2013no, hahn2015construction, yoon201626t, jang2017design, kim2016400}.\\

\par In 2016, an HTS tape manufacturer (SuNAM Co. LTD., Korea) demonstrated the capability of reaching a strong magnetic field of 26.4\,T in a small 35\,mm diameter bore using the HTS technology\,\cite{yoon201626t}. The HTS magnet adopted a No-Insulation (NI) winding method in which the surface of a superconductor tape is not electrically insulated between turns of coil layers. This method allows the overloaded currents induced by unexpected temperature rising to dissipate through the radial direction of the coil. This efficient energy release mechanism enables the self-protection of the magnet from quenching\,\footnote{On 24 December 2020, a sudden dissipation of current from the 18\,T HTS magnet was observed during the scheduled ramp down of the magnet after a month-long detector operation. The observed phenomenon is magnet quenching. The cause of this quenching is not identified, but it is presumed that the current suddenly dropped due to some electrical glitch in the power supply. Most magnet materials were intact, but some magnet coils and insulation parts were damaged due to significant vibrations during rapid energy dissipation. The cavity was slightly bent while all other detector components, including DR, were not affected by the incident.}. Therefore, the HTS NI magnet technology can provide a quench-safe strong magnetic field in a compact volume, which is required for an axion haloscope. A feasibility study of using an HTS magnet for a haloscope experiment is reported in Ref.~\cite{yoon201626t,Ahn_2017}. \\

\par An engineering simulation by SuNAM suggested that an 18\,T and 70\,mm diameter bore solenoid magnet can be reliably built with the existing HTS technology. The 18\,T magnet, if applicable, is the strongest magnet among axion haloscopes. The engineering and technical experience of building such a high field, large-bore HTS magnet may eventually advance the technology of building HTS magnets with much higher magnetic field and larger bore sizes. Therefore, one of the purposes of this experiment is to pioneer magnet technology for future axion haloscopes. FIG.~\ref{fig:18Tmagnet} shows a photograph of the 18\,T HTS solenoid magnet which is designed and built for the axion haloscope experiment. The magnet consists of a stack of 44 double-pancake-coils (DPC) made of $\rm{GdBa_2Cu_3O_{7-x}}$ (GdBCO) tapes~\cite{SuNAM2020a}. All of the coils used in the magnet are no-insulation coils. The peak hoop stress is estimated to be 282\,MPa. Each DPC has two lap joints, and a splice joint is between the DPCs. The total resistance of the magnet is estimated to be 8.1\,$\mu\Omega$ at 4.2\,K, where 4.4\,$\mu\Omega$ is from the lap joints and 3.7\,$\mu\Omega$ is from the splice joints. Multi-width combinations of HTS tapes are used to increase the current density of the magnet. The standard operation current is 199.2\,A at 18.0\,T, and the critical current is 225\,A. The key specifications of the magnet are summarized in Table~\ref{tab:18Tmagnet}, and engineering details can be found in Ref.~\cite{SuNAM2020a}. \\

\begin{table}[b!]
\caption{\label{tab:18Tmagnet} Specifications of the 18\,T HTS magnet}
\begin{ruledtabular}
\begin{tabular}{llrcc}
\multicolumn{2}{l}{Item} &  & \multicolumn{2}{c}{Value}\\
\hline
\multicolumn{2}{l}{Magnetic field}  &  & \multicolumn{2}{l}{18\,T at 199.2\,A}\\
\multicolumn{2}{l}{Stored energy}  &  & \multicolumn{2}{l}{0.93\,MJ}\\
\multicolumn{2}{l}{Field uniformity} &  & \multicolumn{2}{l}{$z=\pm$100\,mm ($>93$\,\% B-field)}\\
\multicolumn{2}{l}{Inner / Outer diameter} &  & \multicolumn{2}{l}{70.0\,mm / 155.6\,mm}\\
\multicolumn{2}{l}{Magnet length} &  & \multicolumn{2}{l}{466.6\,mm }\\
\multicolumn{2}{l}{Total resistance} &  & \multicolumn{2}{l}{0.18814\,$\Omega$}\\
\multicolumn{2}{l}{Total inductance} &  & \multicolumn{2}{l}{18.9\,H}\\
\multicolumn{2}{l}{Ramping rate}  &  & \multicolumn{2}{l}{0.02\,A/s ($<$180\,A) }\\
\multicolumn{2}{l}{} &  & \multicolumn{2}{l}{0.01\,A/s ($>$180\,A) }\\
\multicolumn{2}{l}{Current charging time} &  & \multicolumn{2}{l}{8 hours (0\,A $\rightarrow$ 199.2\,A)}\\
\multicolumn{2}{l}{Field 18\,T charging time} &  & \multicolumn{2}{l}{10 hours (0\,T $\rightarrow$ 18\,T)}\\
\multicolumn{2}{l}{Operating temperature} &  & \multicolumn{2}{l}{4.2\,K (Liquid Helium)}\\
\end{tabular}
\end{ruledtabular}
\end{table}

\par
\begin{figure}[t!]
\centering
\includegraphics[width=0.98\linewidth]{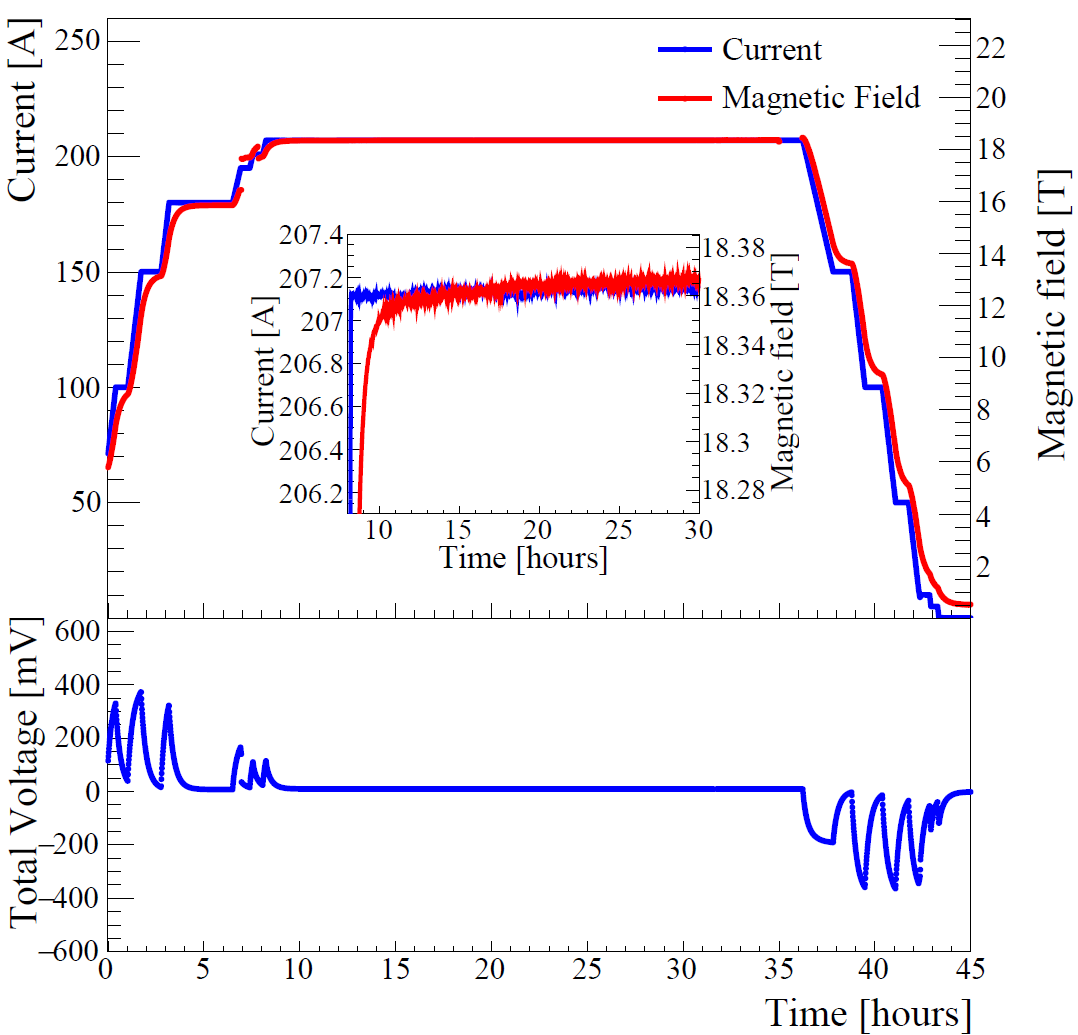}
\caption{First test results of the 18\,T magnet. The top panel shows the measured magnetic field (solid red line) and applied current (solid blue curve). Two missing data regions of the magnetic field at 8 and 36 hours corresponded to when magnetic field mappings were carried out. The inset figure in the top panel compares the applied current and the measured magnetic field. The bottom panel shows the variation of the total magnet voltage.}
\label{fig:18T1stTest}
\end{figure}

 The first test of the 18\,T magnet was carried out in September 2017. For the magnet's continuous and stable operation, the cryostat liquid helium (LHe) level is automatically controlled using feedback loops of the temperature, pressure, and LHe level. FIG.~\ref{fig:18T1stTest} shows the test results. The magnet current is measured using a 0.167\,m$\Omega$ shunt resistor (YOKOGAWA 2215 16, 300\,A, 50\,mV). The current was energized up to 207\,A (18.7\,T) to test the safety margin of the magnet. The charging speed is set to 0.02\,A/s up to 180\,A, and 0.01\,A/s from 180\,A to the final current of 207\,A. The current ramping is paused several times to lower the total voltage below 50\,mV, allowing ample time for the heat dissipation from the DPCs to the surrounding LHe. The dissipation rate is about 6\,mV/min when the applied current is kept constant. The inset figure in the top panel of FIG.~\ref{fig:18T1stTest} shows that it takes about two hours to reach the targeted magnetic field after the final current is set. This delay is due to the effect of the screening currents in the DPC layers. After the magnet is fully charged, the total voltage of the magnet is maintained below 10\,mV as shown in the bottom panel of FIG.~\ref{fig:18T1stTest}. \\

\par The HTS magnet uses resistive elements at DPC joints and current leads. Developing these joints and leads using superconductor (SC) material is incomplete. These are the sources of the Joule heating that cause a large amount of LHe consumption during the magnet operation. Table~\ref{tab:heatload} shows the expected heat load of the resistive components and the expected LHe consumption rates. The estimation does not include the transfer loss of LHe. The measured LHe consumption rate during magnet charging is $\sim$5.8\,L/hr (or $\sim$140\,L/day). The LHe consumption rate reduced to $\sim$5\,L/hr (or $\sim$120\,L/day) during the stable operation of the magnet at 18.7\,T (207\,A). In the initial test, the magnet was operated for about 25\,hours without any magnetic field instability until the scheduled ramp-down. The stability of the magnetic field is measured to be better than 0.05\,\%. \\

\begin{figure}[t!]
\includegraphics[width=0.85\linewidth]{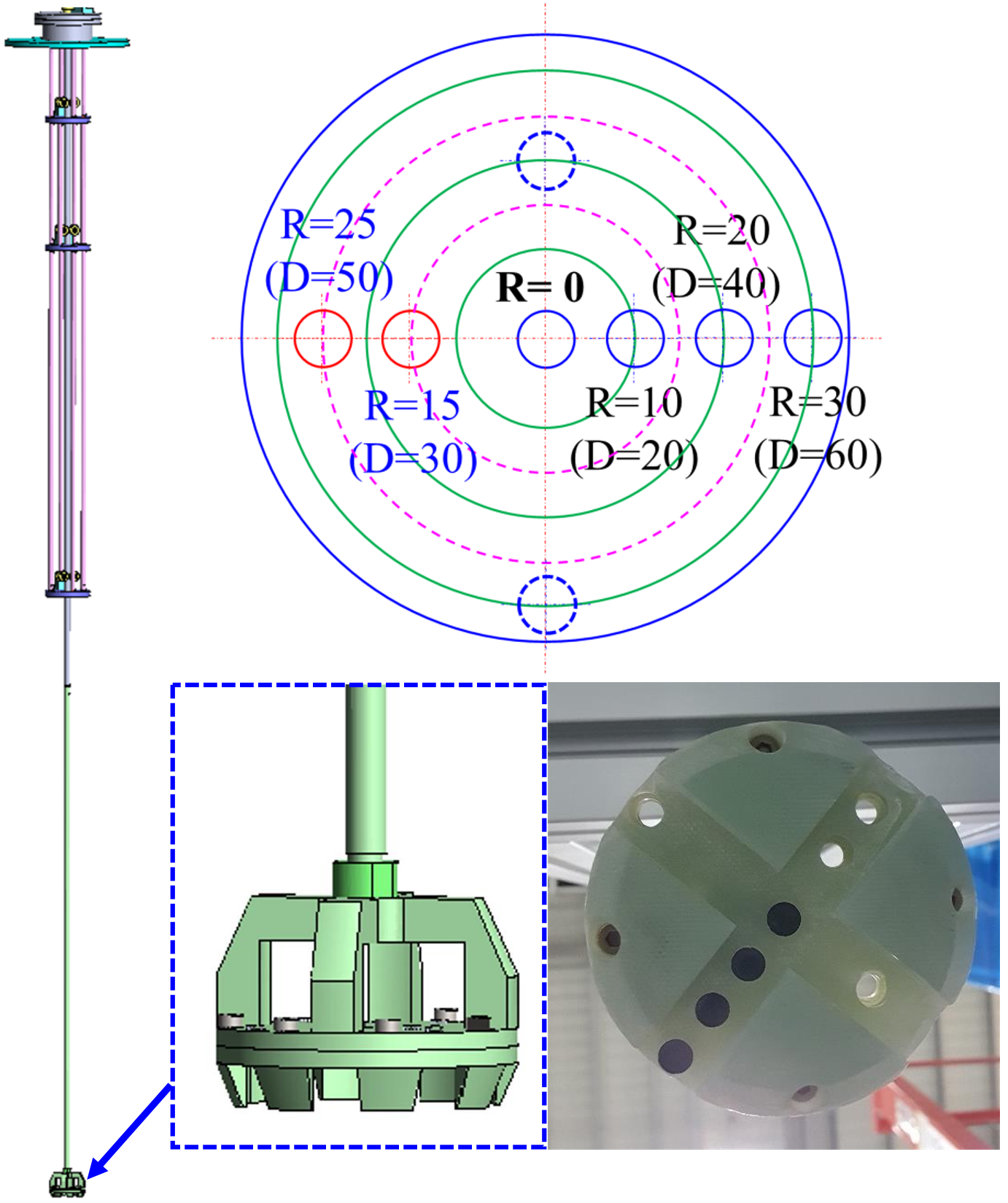}
\caption{The B-field mapper. Left: Side view of the field mapper. Top: Design of the Hall sensor mount. The mount has sensor holders at various locations. Bottom left: Zoom-in view of the Hall sensor mount. Bottom right: A photograph of 4 axial Hall sensors mounted at radial positions of r\,=\,0\,mm, 10\,mm, 20\,mm, and 30\,mm.}
\label{fig:HallSensor}
\end{figure}

\begin{table}[b!]
\caption{\label{tab:heatload} Thermal load of 18\,T magnet components}
\begin{ruledtabular}
\begin{tabular}{lcl}
Heat source & Heat & Remarks \\
\hline
Lap joint            & 0.8\,W    & 200\,n$\Omega$; 88\,ea \\
Splice joint         & 0.8\,W    & 400\,n$\Omega$; 45\,ea \\
HTS current lead     & 0.3\,W    & 4\,$\mu\Omega$, 2 sets \\
Metal current lead   & 0.8\,W    & AWG10, 1.2\,m \\
Single pancake lead  & 0.8\,W    & 6\,$\mu\Omega$, 2 sets \\
Dewar                & 0.4\,W    & \\
\hline 
Total heat load      & 3.9\,W& 5.5\,L/h (1.4\,L/h/W) \\
\end{tabular}
\end{ruledtabular}
\end{table}

\par The axial magnetic fields B$_z(r,z)$ in radial and axial locations are measured using a field probe shown in FIG.~\ref{fig:HallSensor}. Four axial Hall sensors (Lakeshore HGCA-3020) are mounted on a 67\,mm diameter of G10-disk at radial positions of 0\,mm, 10\,mm, 20\,mm and 30\,mm from the center. The G10-disk structure is mounted at the end of a 3\,m-long rod. The supporting structure of the rod is instrumented with pulley systems to guide the vertical movement of the field probe. The probe is inserted into the center of the 18\,T magnet bore. The dynamic range of the probe is from the bottom of the 18\,T magnet to the top of the FCC. \\

\begin{figure}[t!]
\centering
\includegraphics[width=0.98\linewidth]{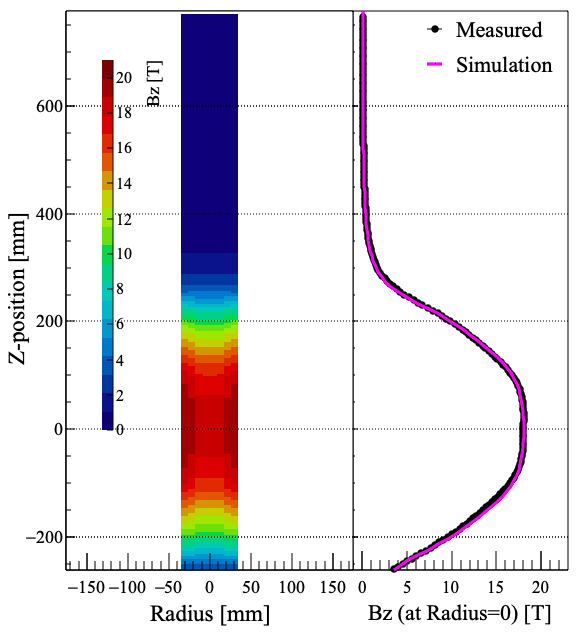}
\caption{Measured axial magnetic fields ($B_z$) of the 18\,T HTS magnet. Left panel shows the radial and z-position dependence of the measured $B_z$ in 2-dimensional projection. Color code indicates the magnetic field strength. Right panel shows the measured (solid dots) and simulated (magenta curve) $B_z$ as a function of the z-position along the axis center (r=0\,mm). Note that $B_z$ at the location of the JPC ($z=+635$\,mm) is measured to be 618\,G.}
\label{fig:18TfieldmapXZ}
\end{figure}

\par FIG.~\ref{fig:18TfieldmapXZ} shows the magnetic field profile of the 18\,T magnet ($B_z$) measured along the magnet bore's parallel axis. $B_z$s are measured from $z$\,=\,-260\,mm (bottom of the magnet) to $z$\,=\,+775\,mm (top of the FCC) for every 5\,mm step. The center of the 18\,T magnet is set to $z=0$\,mm. The uncertainty of the position measurement is about 0.1\,mm. The expected magnetic field is estimated using a simulation package. The small asymmetry in the measured field may have been caused by the magnetization of the stainless steel ring at the top of the magnet\,(See FIG.~\ref{fig:18Tmagnet}). The off-center ($r>0$\,mm) measurements show that $B_z$s are higher at the cylinder wall; for example, at the magnet center ($z=0$\,mm), the measured magnetic fields are 18.03\,T at $r$=0, 18.06\,T at $r$=10\,mm, 19.02\,T at $r$=20\,mm, and 19.31\,T at $r$=30\,mm. \\

\subsection{Dilution Refrigerator}
\par A wet-type DR, Kelvinox-400 by Oxford Instruments, is used to cool down the detector. The cooling power of the DR is 400\,$\mu$W at the mixing chamber temperature of 120\,mK. The DR uses a Joule-Thomson (JT) heat exchanger. The cold plate (Cold) temperature is about 200\,mK, and the Still temperature is about 800 to 900\,mK. The base temperature at the mixing chamber without any thermal load is about 30\,mK. In addition to the preinstalled thermometers in the DR, four additional calibrated Ruthenium Oxide (RuO$_2$) thermometers are installed at the mixing chamber, JPC mount, cavity, and detector mount frame. These temperatures are periodically monitored during the detector operation. The detector mount frame is constructed with Oxygen-Free-High-Conductivity (OFHC) copper structure and aluminum plates and is thermally anchored on the mixing chamber. The JPC mount structure is thermally anchored at the top plate of the detector frame using a gold-plated copper jig. The cavity is mounted at the bottom of the detector frame. It takes a few hours to cool down the detector system from 4\,K to 60\,mK. A typical temperature of the mixing chamber during a regular detector operation is about 50$\sim$60\,mK, the JPC temperature is about 60\,mK, and the cavity temperature is about 100$\sim$120\,mK.\\

\subsection{Microwave Cavity and Frequency Tuning}
\begin{figure}[t!]
\centering
\includegraphics[width=0.10\linewidth]{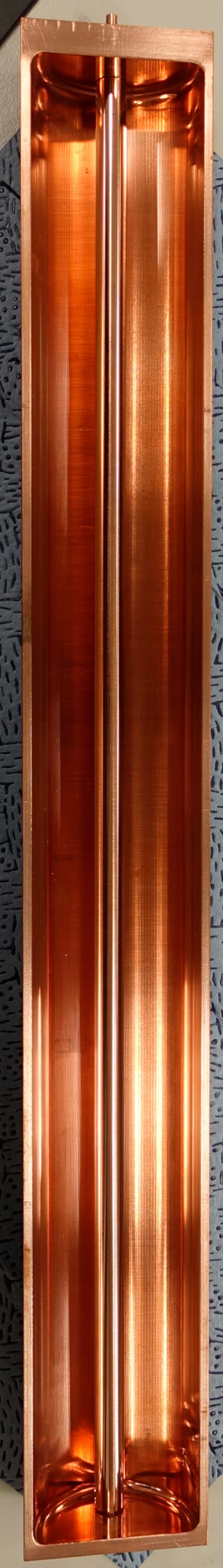} 
\includegraphics[width=0.65\linewidth]{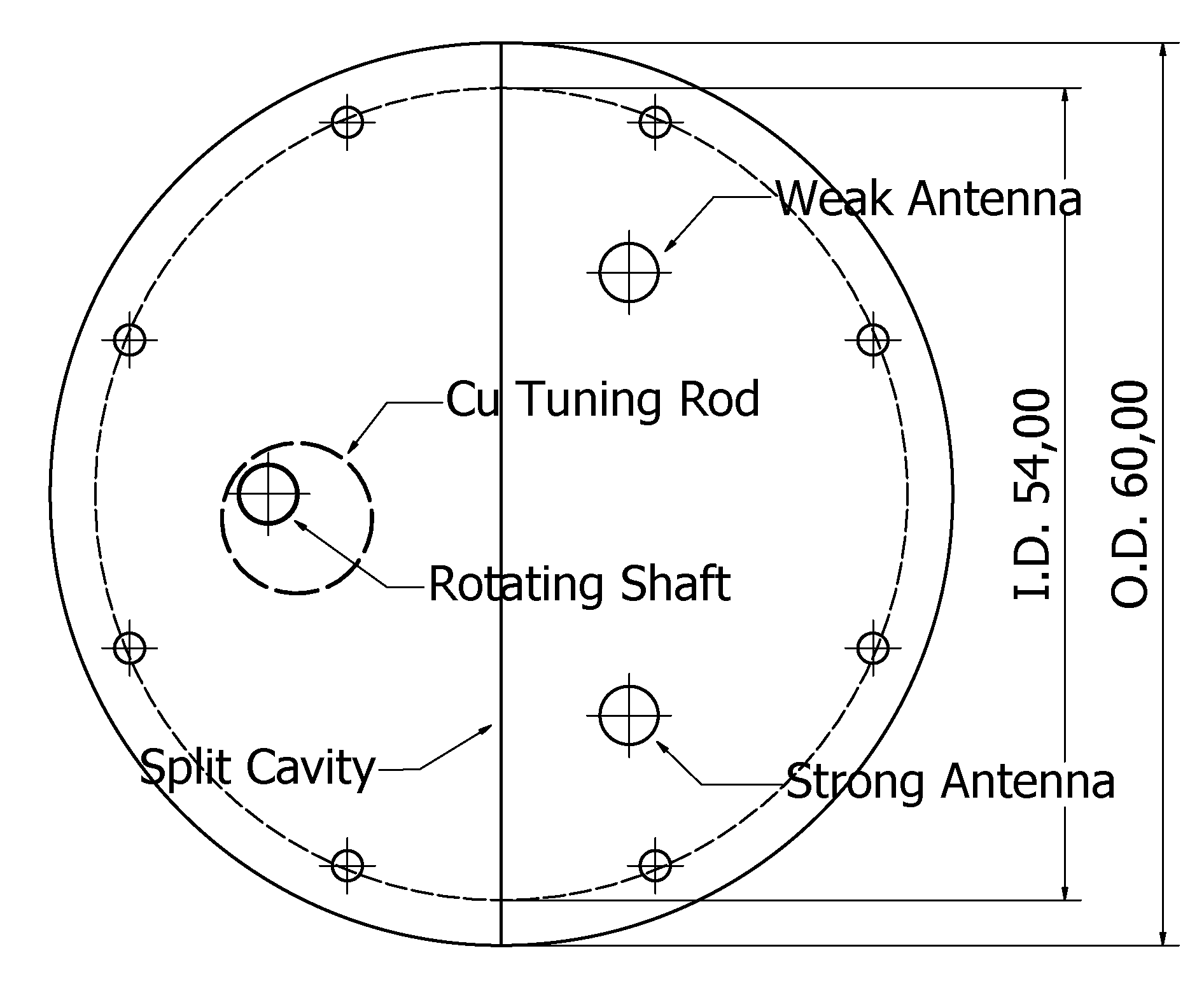}
\caption{Microwave cavity. Left: Photograph of a vertically spilt cavity with a tuning rod. Right: Top view drawing of the cavity which shows the locations of the tuning rod and two RF antennas.}
\label{fig:cavity}
\end{figure}

\par Electromagnetic waves in the resonant cavity are categorized into three distinct types: transverse electric mode (TE, $E^{\prime}_z$\,=\,0), transverse magnetic mode (TM, $B^{\prime}_z$\,=\,0), and transverse electromagnetic mode (TEM, $E^{\prime}_z$ and $B^{\prime}_z$). The form factor of the resonant cavity in Eqn.\,\ref{eqn:pa} is defined as
\begin{equation}
C_{nlm}=\frac{{\left|\int_V{\bold{E^{\prime}}_{nlm}\cdot\bold{B_0}}\,dV\right|^2}}{\int_V{\left|\bold{E^{\prime}}_{nlm}\right|^2dV}\cdot\int_V{\left|\bold{B_0}\right|^2dV}}. 
\end{equation}
\noindent In a solenoid, the magnetic field is applied parallel to the cavity axis. Therefore, the form factor is determined by the axial component of the electric field of the cavity mode. In TE and TEM modes, the form factor becomes zero. Accordingly the signal power in Eqn.\,\ref{eqn:pa} vanishes. Therefore, only the TM modes result in non-zero signal power. The axial electric field of the TM mode in an empty cylindrical waveguide is expressed as
\begin{equation}
E_z (r ,\phi ,z)=E(t)J_n \left( \frac{x_{nl}}{R} r \right) e^{\pm in \phi} \cos \left( \frac{m \pi z}{L} \right),
\end{equation}
\noindent where $E(t)$ is the electric field, $J_n$ is the $n^{\text{th}}$ Bessel function, $x_{nl}$ is the $l^{\text{th}}$ solution of $J_n (x)=0$, $R$ is the inner radius of the cavity, and $L$ is the cavity height. The form factor is maximum at TM$_{010}$ mode, and therefore, the signal power is  maximized as well. The resonant frequency of the TM mode is determined as

\begin{equation}
{\nu}_{nlm} = \frac{1}{2 \pi \sqrt{{\mu}_0 {\epsilon}_0}} \sqrt{{\left( \frac{x_{nl}}{R} \right)}^2 + {\left( \frac{m \pi}{L} \right)}^2},
\end{equation}
\noindent where $\mu_0$ ($\epsilon_0$) is the permeability (permittivity) of vacuum. At the TM$_{010}$ mode, the resonant frequency is determined by the radius of the cavity.  \\

\begin{figure}[t!]
\centering
\includegraphics[width=1.\linewidth]{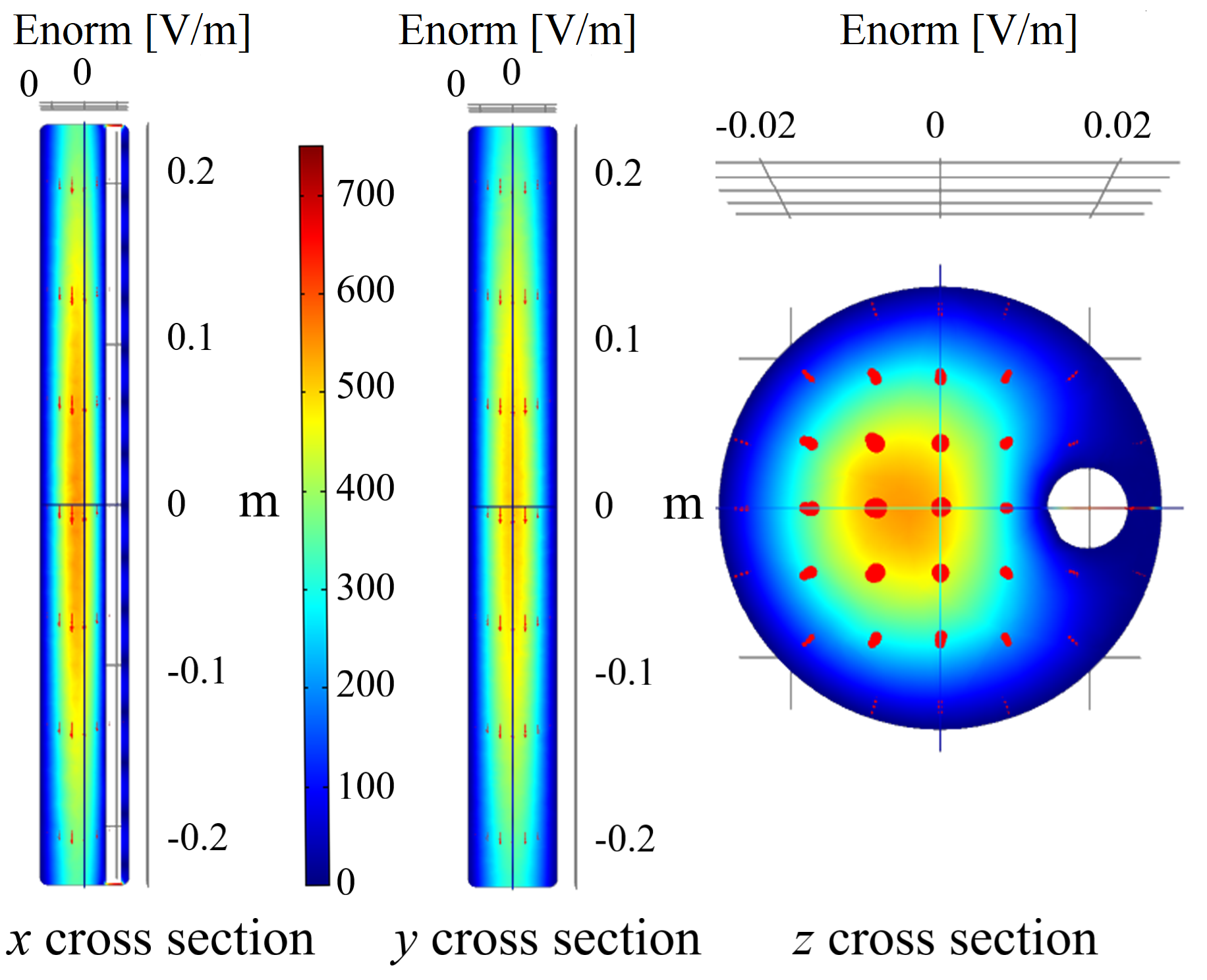}
\caption{Electric field of the TM$_{\rm{010}}$ mode at 4.744\,GHz. The left figure is a section view in the x-plane. The tuning rod is shown as a vertical white bar. The middle figure is a section view in the y-plane. Red arrows in the right figure are the electric field vectors.}
\label{fig:Enorm}
\end{figure}

\par The cavity, made of an OFHC copper, is vertically split into two half-cylinder pieces to suppress the induced eddy current during the magnet ramping. These half-cylinders are hollowed out, with 54.0\,mm in diameter and 466.6\,mm in length. The split surface of the half-cylinder is electrically insulated using Kapton tapes. The two half-cylinders are tied with copper flanges at the top and bottom of the cavity. The empty volume of the cavity is 1.07\,L. The resonant frequency of the cavity ($\nu_C$) is tuned using a cylindrical OFHC copper tuning rod. The dimension of the rod is 10.0\,mm in diameter and 465.6\,mm in length. 
The tuning rod shaft is 2.5\,mm off the cavity center. The top shaft of the tuning rod is made of PEEK to prevent RF power leakage from the cavity. The bottom shaft of the tuning rod is made of an OFHC copper, which is thermally coupled to the pivot structure at the cavity bottom. FIG.~\ref{fig:cavity} shows a photograph of the half-cavity in which the tuning rod is installed. The drawing at the right shows the locations of the tuning rod and two RF antennas. The locations of the two antennas are determined to minimize the interference of the RF signals. Due to the broken geometrical symmetry by the tuning rod, radial electric fields are formed between the tuning rod and the inner surface of the cavity. These radial electric fields reduce the form factor\,\cite{Hagmann1990a, Hagmann1990b}. FIG.~\ref{fig:Enorm} shows the electric field distribution of the TM$_{\rm{010}}$ mode at the resonant frequency of 4.744\,GHz. The red arrows represent the electric field vectors that vanish at the cavity's top and bottom. The inconstancy of the electric field along the z-axis is caused by the tuning rod and the fillets at the top and bottom of the cavity.\\

\begin{figure}[t!]
\centering
\includegraphics[width=1.0\linewidth]{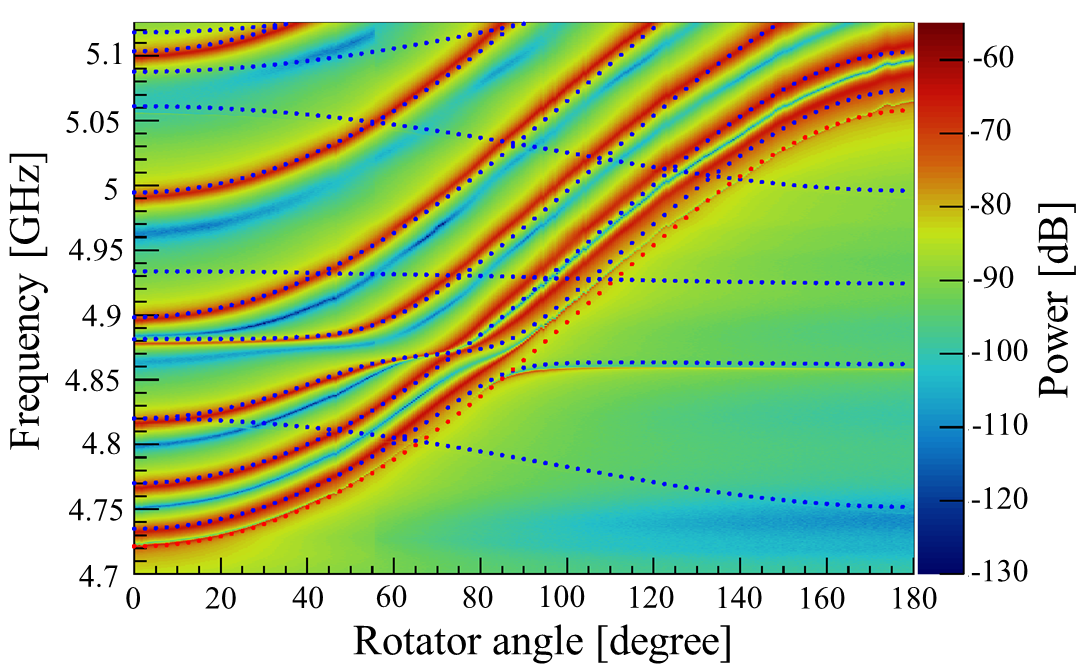}
\caption{Measured and simulated mode maps of a cavity at RT. The color code represents the measured EM power in dB. Simulations are shown in dotted points. TM$_{\rm{010}}$ mode is shown in red dots, while other modes are shown in blue. The tuning rod is farthest to the cavity center at 0 degrees and closest at 180 degrees. The TM$_{\rm{010}}$ mode crosses four other EM modes. Among them, two crossing modes at 4.802\,GHz and 4.852\,GHz are significant enough to be visible in this measured mode map.}
\label{fig:Modemap}
\end{figure}

\par FIG.~\ref{fig:Modemap} shows the measured and simulated resonant EM modes in RT as a function of the rotation angle of the tuning rod. The TM$_{\rm{010}}$ mode covers the frequency range of 4.72\,GHz to 5.06\,GHz. Four mode crossings are found in this frequency range, where the EM modes become dubious. In cryogenic temperature (CT), the lowest crossing mode is found at 4.840\,GHz, and the higher crossing modes are at 4.879\,GHz, 4.948\,GHz, and 5.042\,GHz. None of the crossing modes resides within the search region of the first phase axion dark matter experiment. \\

\par The tuning rod and strong coupling antenna positions are mechanically moved to tune $\nu_C$ and $\beta$. Three micro stepper motors (Oriental Motor PKE564AC-HS100) in RT are used to precisely control the mechanical motion. The shaft of the RT motor is coupled to the rotary feedthrough of the vacuum joint. A long G10-rod connects the RT stage joint to the 4K stage pulley for thermal insulation. Both ends of the G10-rod are connected by 0.36\,mm thick Kevlar threads. Kevlar threads are good thermal insulators with thermal conductivity of 0.04\,W/m$\cdot$K. A downstream Kevlar thread is guided to an 85\,mm diameter brass disk to drive a right-handed rotation of the tuning rod. The Cryogenic bearing hub for the disk axis is mounted on an aluminum plate on the detector frame. The axis of the disk is mechanically coupled to the tuning rod shaft. The Kevlar thread end is tied to the side of the disk so that pulling the Kevlar thread rotates the disk in the right-handed direction. The torsion spring on the disk axis provides the recovery force for a backward rotation. In addition to the torsion spring, a supplementary Kevlar-pulley system provides a forced backward rotation. Another stepper motor controls the linear motion of the strong antenna. A Kevlar thread guides the linear motion of the strong antenna, and a vertical spring provides the recovery force when the antenna is pulled upward. The resolution of the linear motion is about 0.6$\mu$m per 10$^3$ steps. The weak antenna is fixed to the optimal position, as the fine-tuning of the weak antenna position did not significantly affect the tuning parameters. \\

\par The unloaded quality factor of the cavity is given as $Q_0 = (\mu_0/\mu_C)(V/A\delta)\xi$, where $\mu_C$ is the permeability of the inner wall of the cavity, $A$ is the surface area, $\delta$ is the skin depth, and $\xi$ is the geometry factor\,\cite{jackson1999classical}. In RT, the resistivity of metals increases under the magnetic field. However, in CT with increasing magnetic fields, conduction electrons are confined in smaller orbits in the surface area. This anomalous effect reduces resistivity and decreases the skin depth\,\cite{Ahn_2017}. As a result, under this condition, $Q_0$ and $Q_L$ increase. In the current setup, within the range of the scanned frequency, $Q_L$ at the critical coupling ($\beta \simeq 1$) is about 6\,000 to 9\,000 at RT and 14\,000 to 29\,000 at CT, depending on $\nu_C$.

\subsection{Cavity Field Profile and Form Factor Uncertainty}
\par The EM field profile in the resonant cavity can be probed by a {\it bead perturbation method}~\cite{Rapidis2019}.  In this method, a small dielectric bead is placed in the cavity where the electric field variation is the largest. The frequency shift by the field perturbation is given by
\begin{eqnarray}
\frac{\omega-\omega_0}{\omega}=-\frac{\int_{V_0}(\Delta\epsilon\vec{E}\cdot\vec{E_0}+\Delta\mu\vec{H}\cdot\vec{H_0})dV}{\int_{V_0}(\epsilon\vec{E}\cdot\vec{E_0}+\mu\vec{H}\cdot\vec{H_0})dV},
\label{eqn:perturb01}
\end{eqnarray}
\noindent where $\omega_0$ is the unperturbed resonant frequency, $\vec{E_0}$ and $\vec{H_0}$ are the unperturbed electric and magnetic fields, $\vec{E}$ and $\vec{H}$ are the perturbed electric and magnetic fields, $\epsilon$ ($\mu$) is the permittivity (permeability) of an unperturbed cavity, and $\Delta\epsilon$ ($\Delta\mu$) is the permittivity (permeability) difference between the perturbed and the unperturbed cavity\,\cite{Pozar2012}. Assuming that $\Delta\epsilon$ and $\Delta\mu$ are small, $\vec{E}$, $\vec{H}$, and $\omega$ in the denominator can be approximated to $\vec{E_0}$, $\vec{H_0}$, and $\omega_0$, respectively. As the volume of the bead is small compared to that of the cavity, Eqn.\,~\ref{eqn:perturb01} can be approximated as
\begin{eqnarray}
\frac{\Delta \omega}{\omega_0}\simeq-\frac{(\epsilon_b-1)V_\text{b}\,E(\mathbf{r})^2}{W},
\label{eqn:BeadPert01}
\end{eqnarray}
\noindent where $\Delta \omega = \omega - \omega_0$, $\epsilon_b$ is the dielectric constant of the bead, $V_\text{b}$ is the volume of the bead, and $W$ is the stored energy inside the cavity. Therefore, the electric field at the bead location can be estimated by measuring the fractional shift of the resonant frequency. \\

\begin{figure}[t!]
	\centering
	\includegraphics[width=1.\linewidth]{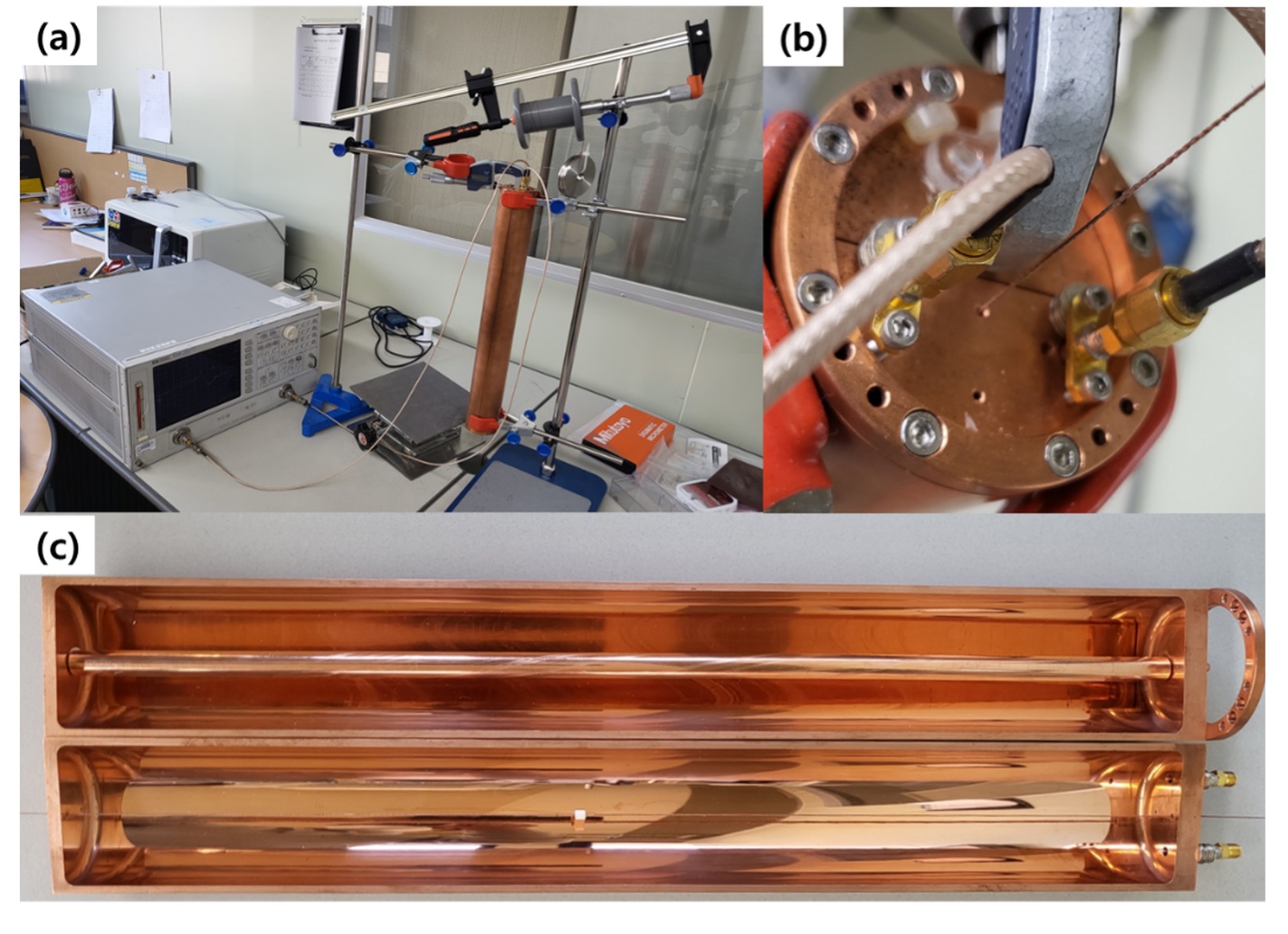}
	\caption{(a) Overall test setup for the bead perturbation measurement. (b) Top view of the cavity. Two antennas are inserted into the cavity top. (c) Half split view of the cavity. The copper tuning rod is placed in a half-cylinder. The alumina bead is shown in the middle of the other half-cylinder. }
	\label{fig:BP_setup}
\end{figure}

\begin{figure*}
\centering
\includegraphics[width=1.\linewidth]{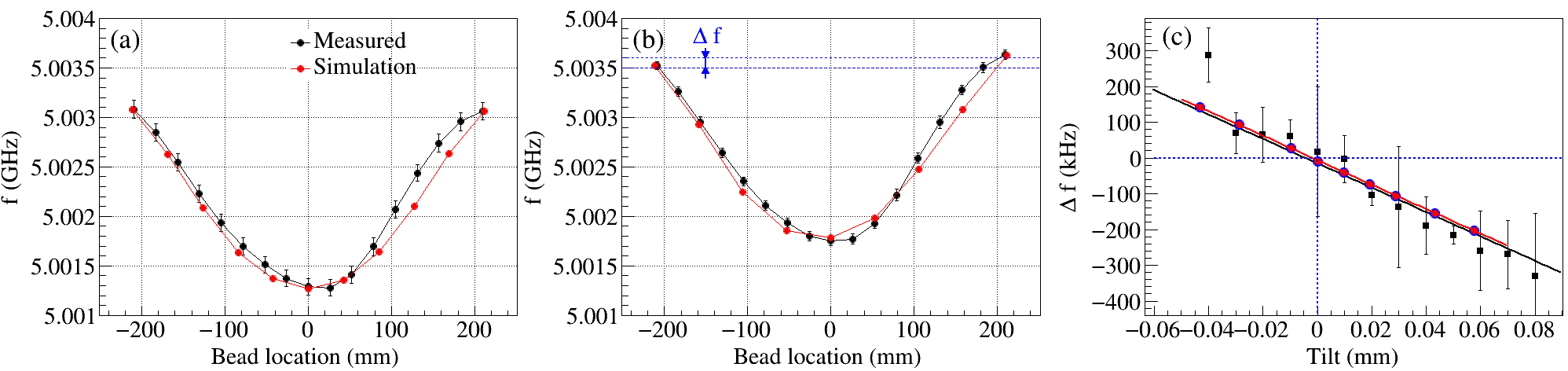}
\caption{Results of the bead perturbation measurements. The measured data (black dot) and simulation results (red dot) are overlayed. (a) An example of the perturbed resonant frequencies when the tuning rod is aligned along the z-axis. (b) An example of the resonant frequencies when the tuning rod is tilted by 0.02\,mm offset at the top, which causes non-zero $\Delta f$. (c) $\Delta f$ as a function of the tilt offset.}
\label{fig:Df_tilt}
\end{figure*}

\par The test setup for the bead perturbation measurement is shown in FIG.~\ref{fig:BP_setup}. The cavity cylinder consists of two half-vertical pieces. The bead moves in the axial direction in one half, while a tuning rod is placed in the other half. The bead is a section of alumina ($\epsilon_b=9.8$) tube with a height of 4.05\,mm, an outer diameter of 3.80\,mm, and an inner diameter of 0.95\,mm. A Kevlar thread through the bead bore is knotted at the bead. The Kevlar line passes through the top and bottom holes of the cavity. The diameter of the Kevlar thread is 0.5\,mm and the diameter of the holes at the top and bottom of the cavity is 1.0\,mm. The hole is located 5.0\,mm from the center at the top and the bottom sides. The Kevlar line is wound around a pulley to control the bead position. The bead can move along the line parallel to the z-axis at constant radial and azimuthal positions where the variation of the electric field is maximal. A VNA measures the frequencies of the TM$_{010}$ modes.\\

\par The asymmetries of the cavity configuration, such as the tilt of the tuning rod and the uneven gap between the split space, result in field localization and form factor degradation\,\cite{Hagmann1990b}. A micrometer is mounted on the cavity top to configure the asymmetric geometry by the tuning rod. The tilt angle of the tuning rod is controlled by moving the top axle by 0.01\,mm per step. The deformation of the TM$_{010}$ mode is measured by moving the bead position. The field localization from the tuning rod tilt is parametrized by the resonance frequency asymmetry ($\Delta f$) between the top and the bottom bead positions. The $\Delta f$ is zero if the tuning rod is perfectly parallel to the cavity and becomes non-zero as the tuning rod is tilted. The measurements are repeated 15 times at each tilting offset position. Simulation studies are conducted using the COMSOL Multiphysics\textsuperscript{\textregistered} package. The results are compared with the data shown in FIG.~\ref{fig:Df_tilt}. A linear correlation between the $\Delta f$ and the tilt offset is observed as expected in the simulation study. The gaps between the tuning rod shafts and the shaft holes on the cavity's top and bottom are 0.35\,mm. However, the mechanically allowed tilt offset of the full detector setup in DR is about 0.1\,mm due to the mount structures of the tuning rod axle. \\

\par The form factor uncertainty due to the field asymmetry is evaluated using the simulator. The form factor reduction solely by the 0.1\,mm of tuning-rod tilt is 1.9\,\%. The measured uneven split gap difference between the top and the bottom gaps is less than 0.05\,mm in RT. However, we allow the potential uneven gap difference of 0.1\,mm in CT to be conservative. The form factor degradation caused solely by the uneven gap is 0.47\,\%. The mechanical tolerance of the cavity fabrication is better than 0.01\,mm in radius, and the systematic uncertainty due to the fabrication error is negligible compared to the other uncertainties above. The total form factor uncertainty is evaluated under the combined asymmetry conditions of both the tuning rod tilt and the uneven split gap. A simulation study is carried out on ranges beyond the limit of the tilt and gap differences to study the overall tendency of the form factor degradation. Through this study, the form factor uncertainty is conservatively estimated as 3.9\%.

\subsection{Josephson Parametric Converter}
\begin{figure}[t!]
\centering
\includegraphics[width=0.5\linewidth]{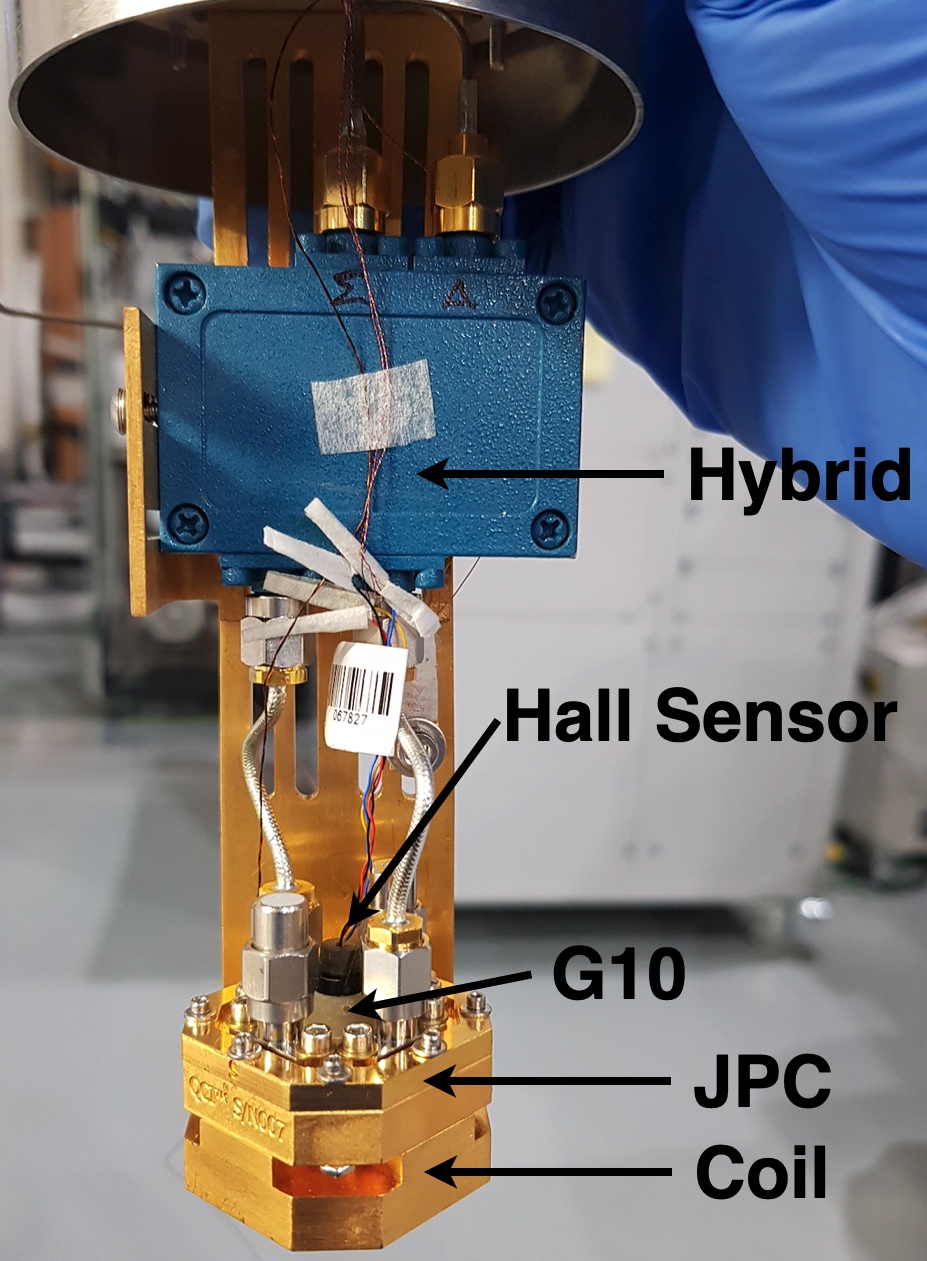}
\caption{Photograph of the JPC unit. The JPC in this setup is wired on the idler mode. A 180$^\circ$-Hybrid circuit is connected for a proper excitation. A flux coil for the frequency control is installed under the JPC casing. The Hall sensor is mounted about 1\,cm above the JPC casing on a G10-base to dilute Joule heating. To reduce Joule heating from the Hall sensor operation, the sensor's current is set to 4\,mA. This whole structure is inserted into double layers (Amuneal and aluminum) of passive shield cans.}
\label{fig:JPC}
\end{figure}

\par Quantum-noise-limited RF amplifiers are an essential component of the modern axion haloscope experiments for high-fidelity measurements. A JPC is used as the first stage RF amplifier. The JPC is a phase-preserving superconducting parametric amplifier from Quantum Circuits, Inc. In a phase-preserving amplifier, the gains of the in-phase and out-of-phase quadratures are the same, while in a phase-sensitive amplifier, the gains of the quadratures are reciprocals. In a parametric amplifier, an appropriate pump tone drives the nonlinearity in the inductance of the circuit, which amplifies the system. The JPC circuit is formed with four identical Josephson junctions consisting of a superconducting loop. The resonant frequency can be tuned by changing the magnetic flux through the ring-modulator\,\cite{bergeal2010phase, Bergeal2010JPC, LiuJPC2017}. \\

\par FIG.~\ref{fig:JPC} shows our JPC unit together with the associated components. An external RF pump tone powers the JPC. The JPC has two amplifying modes called signal and idler modes. The signal mode is in the frequency range of 7.720\,GHz to 8.802\,GHz ($\Delta f$=1\,082\,MHz), and the idler mode is in the range of 4.757\,GHz to 5.010\,GHz ($\Delta f$=253\,MHz). The Idler mode is used in the 1st phase experiment. The bias current on the coil controls the magnetic flux through the JPC modulator, which changes the critical current of the JPC. Accordingly, the value of the inductance changes along with the resonant frequency of the JPC ($\nu_J$). While $\nu_J$ mainly depends on the bias current of the coil, the power and frequency of the pump tone slightly shift $\nu_J$ as well. The power of the pump tone tunes the peak gain of the JPC ($G_J$). \\

\begin{figure}[t!]
\centering
\includegraphics[width=0.98\linewidth]{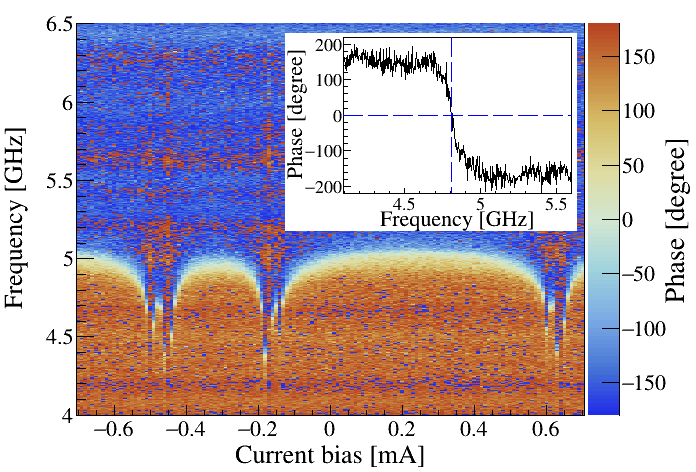}
\caption{Frequency-phase diagram as a function of bias current. Color code represents the phase of JPC. The inset plot is the section view of the phase transition at the bias current of -0.12\,mA.}
\label{fig:JPCphasemap}
\end{figure}

\par
\begin{figure}[b!]
\centering
\includegraphics[width=0.98\linewidth]{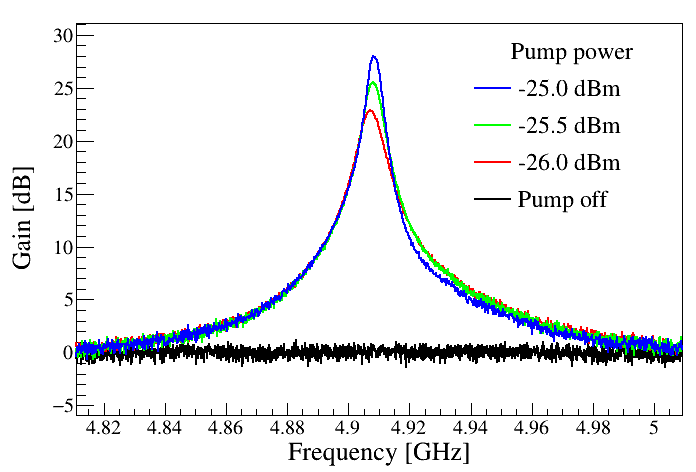}
\caption{JPC gain measurement for three different pump power setups. Black trace shows the normalized background noise at 0\,dB which is measured when the JPC pump is off. In this particular example, the pump frequency is fixed at 13.515\,GHz while varying the pump power.}
\label{fig:JPCgain}
\end{figure}

 FIG.~\ref{fig:JPCphasemap} shows the measured phase responses of the JPC idler mode as a function of the bias current. The inset plot shows a section view of the JPC phase response, in which the phase changes 2$\pi$ at the resonance frequency. There are large and small lobes in the frequency-phase diagram. The amplification frequency range is chosen from 4.7\,GHz to 5.0\,GHz, which corresponds to the edges of the large lobe between -0.14\,mA and 0.0\,mA of the bias current. The background noise profile of the JPC is measured without pump tone, which makes the JPC a passive mirror. The JPC pump frequency is set to the sum of the resonance frequencies of the two modes. The JPC pump power is initially set to -30\,dBm, then slowly increases in 0.01\,dBm steps until the target peak gain is achieved. The net gain of the JPC is obtained by dividing the signal response profile by the background noise profile. FIG.~\ref{fig:JPCgain} shows the net gain profile of our JPC for different pump powers. In the axion search operation, the typical peak gain of the JPC is set to 27\,dB. \\

\begin{figure*}[t!]
\centering
\includegraphics[width=0.35\linewidth]{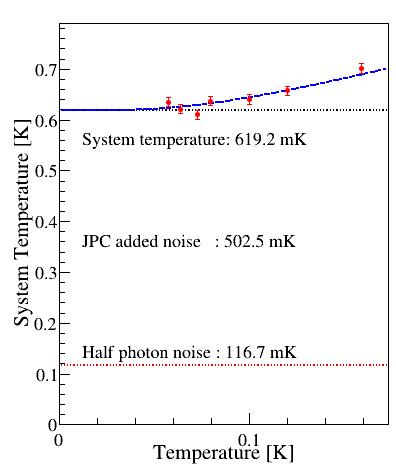} 
\includegraphics[width=0.35\linewidth]{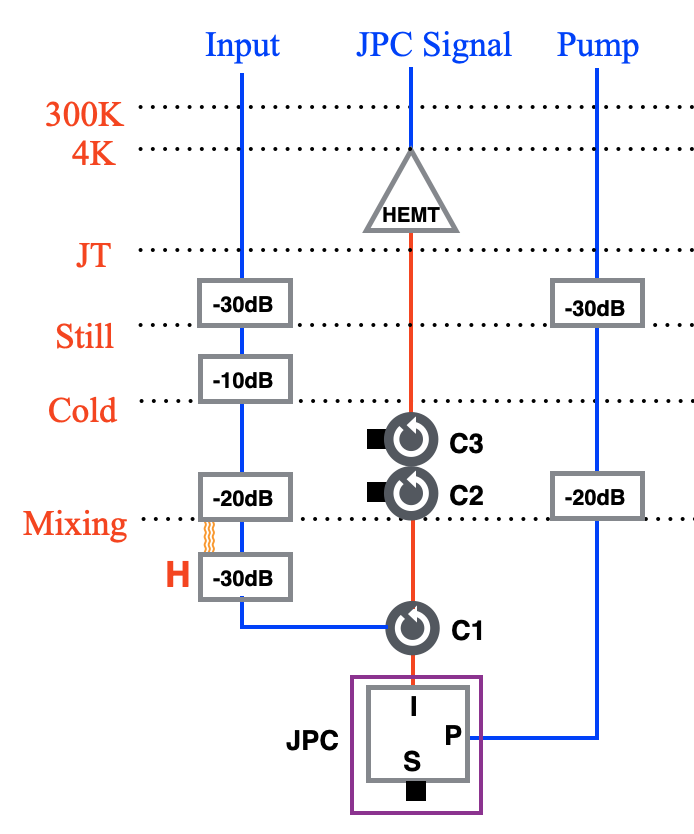}
\caption{JPC noise temperature measurement. The schematic diagram on the right shows the RF chain layout. The heater (H) is made of a 30\,dB attenuator wrapped with a copper heater block. A calibrated RuO$_2$ thermometer is mounted on the heater block. The figure on the left shows the results of the measurement. The data points (red dots with error bars) are the measured noise temperatures. The blue curve shows the expected noise temperature, including the 1/2 photon contribution at 4.9GHz.}
\label{fig:JPCaddednoise}
\end{figure*}

\par The added noise temperature of the JPC is measured with a dedicated CT-RF setup, shown in the schematic diagram in FIG.~\ref{fig:JPCaddednoise}. A heater (H) mounted on a 30\,dB attenuator is thermally coupled to the mixing chamber, using three bundles of copper wires. The input RF noise to the JPC is changed by tuning the physical temperature of the heater. The thermal anchoring of the heater to the mixing chamber is weak enough. The base temperature of the JPC remains at $\sim$60\,mK while the heater's physical temperature varies between 80 to 200\,mK. FIG.~\ref{fig:JPCaddednoise} shows the measured noise temperature as a function of the effective input noise temperature. The total system noise temperature is estimated to be 619.2\,mK. Considering the vacuum noise of a half-photon (116.7\,mK at 4.9\,GHz), the added noise by the JPC is 502.5$\pm$3.4\,mK, which is consistent with the specification of the JPC from the vendor and other measurements\,\cite{Roch2012,Bergeal2012,Spietz2010}. This measured JPC noise temperature is higher than the ideal case. The high noise could be due to the attenuation between the Josephson junctions and HEMT, mismatching of the impedance of the junction\,\cite{Roch2012,Spietz2010}, and the correlation between the gain and noise temperature due to the imperfect isolation of the circulators\,\cite{Bergeal2012}.\\

\subsection{Field Cancellation Coil}
\par
\begin{figure}[t!]
\centering
\includegraphics[width=1\linewidth]{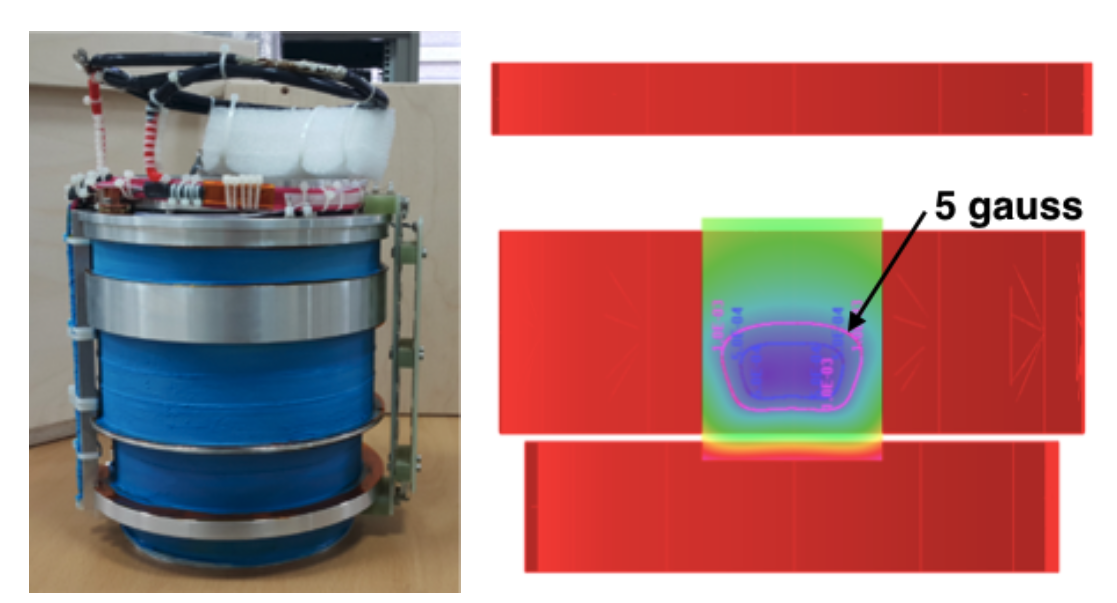}
\caption{Left: A photograph of the FCC. Right: A result of the field simulation by Kiswire Advanced Technology.}
\label{fig:fcc0}
\end{figure}

 The JPC is relatively insensitive to weak ambient magnetic fields. According to the dedicated magnetic field tests, the JPC is stable in about 20\,G of the axial magnetic field, and no significant performance change is observed. However, as shown in FIG.~\ref{fig:18TfieldmapXZ}, the stray field $B_z$ at the location of the JPC ($z$=+$635$\,mm) is measured to be 618\,G when the 18\,T magnet is fully charged. The Superconducting-quantum-interference-device(SQUID) and the ring-modulator in the JPC require a substantially reduced magnetic field environment. Therefore, the stray magnetic field must be maintained below $\sim$20\,G. An FCC is designed and developed to cancel out the stray field. The FCC is built with the LTS wires by Kiswire Advanced Technology, a Korean LTS wire manufacturer. A simulation study for the FCC is carried out to achieve a magnetic field below 5\,G in 50\,mm of the axial range at the center of the FCC. The maximum allowed external field is 1\,400\,G. FIG.~\ref{fig:fcc0} shows a photograph of the FCC and a simulation result of the magnetic field. The FCC consists of three separate LTS solenoids using multi-filamentary NbTi wires to produce a uniform field in the center of the FCC bore. The overall height of the FCC is 246\,mm, including the supporting structure. The bore size of the FCC is 145\,mm in diameter. The specifications of the FCC are listed in table~\ref{tab:fcc}. \\

\begin{table}[b!]
\caption{\label{tab:fcc} Dimensions of the FCC. The FCC consists of three separate LTS solenoids. The width of each coil, inner diameter (ID), outer diameter (OD), and total length of NbTi wire of each coil are listed below.}
\begin{ruledtabular}
\begin{tabular}{c|cccc}
Coil & Height [mm]& ID [mm] & OD [mm] & Wire length [m]\\
\hline
Top    &  28 & 145 & 155 & 65.973 \\
Center & 110 & 159 & 161 & 55.292 \\
Bottom &  30 & 145 & 147 & 13.760 \\
\end{tabular}
\end{ruledtabular}
\end{table}

\par The ratio of the magnetic field to the applied current at the center of the FCC is 16.0\,G/A. The FCC needs to be charged to -38.6\,A to cancel out the stray field of 618\,G. A persistent heater switch of the FCC is powered at 16.0\,V and 0.306\,A. The heater is mainly used to apply or relieve currents to the FCC since the heater breaks the superconductivity of the FCC. As described in the previous section, the Hall sensor is installed above the JPC inside the passive shielding cans. The bias current is set to 4\,mA to reduce Joule heating from the Hall sensor. The Hall sensor was calibrated at the laboratory for Advanced Materials and Extreme Conditions of POSTECH, Korea. \\

\begin{figure}[t!]
\centering
\includegraphics[width=0.98\linewidth]{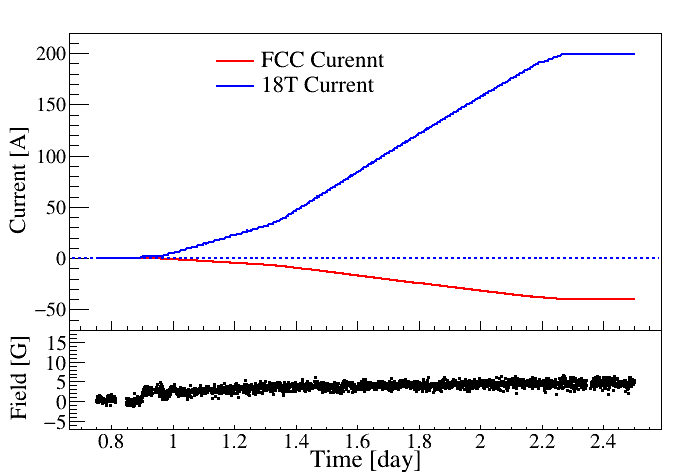}
\caption{Result of dynamic field cancellation during the 18\,T magnet ramping process. The upper panel shows the applied currents on the two magnets. Signs of currents are assigned to indicate the direction of the axial magnetic field. The lower panel shows the measured magnetic field near the JPC. The magnetic field was controlled under 6.5\,G during the test.}
\label{fig:dcancel}
\end{figure}

\par Stray fields need to be canceled out during the 18\,T magnet ramping process to avoid trapping the magnetic flux in the JPC circuit. Therefore, the FCC is charged with the 18\,T magnet to continuously cancel the rising stray fields. FIG.~\ref{fig:dcancel} shows an example of such dynamic field cancellation. Initially, a few Amps of small current are applied to the 18\,T magnet. The magnetic field is then monitored at the location of the JPC. Stray magnetic fields can be suppressed down to near 0\,G level by applying the corresponding cancellation currents to the FCC. The applied currents on the 18\,T magnet, and FCC are simultaneously raised. During the test, the residual magnetic field at the Hall sensor near the JPC was controlled under 6.5\,G. \\

\begin{figure}[t!]
\centering
\includegraphics[width=0.98\linewidth]{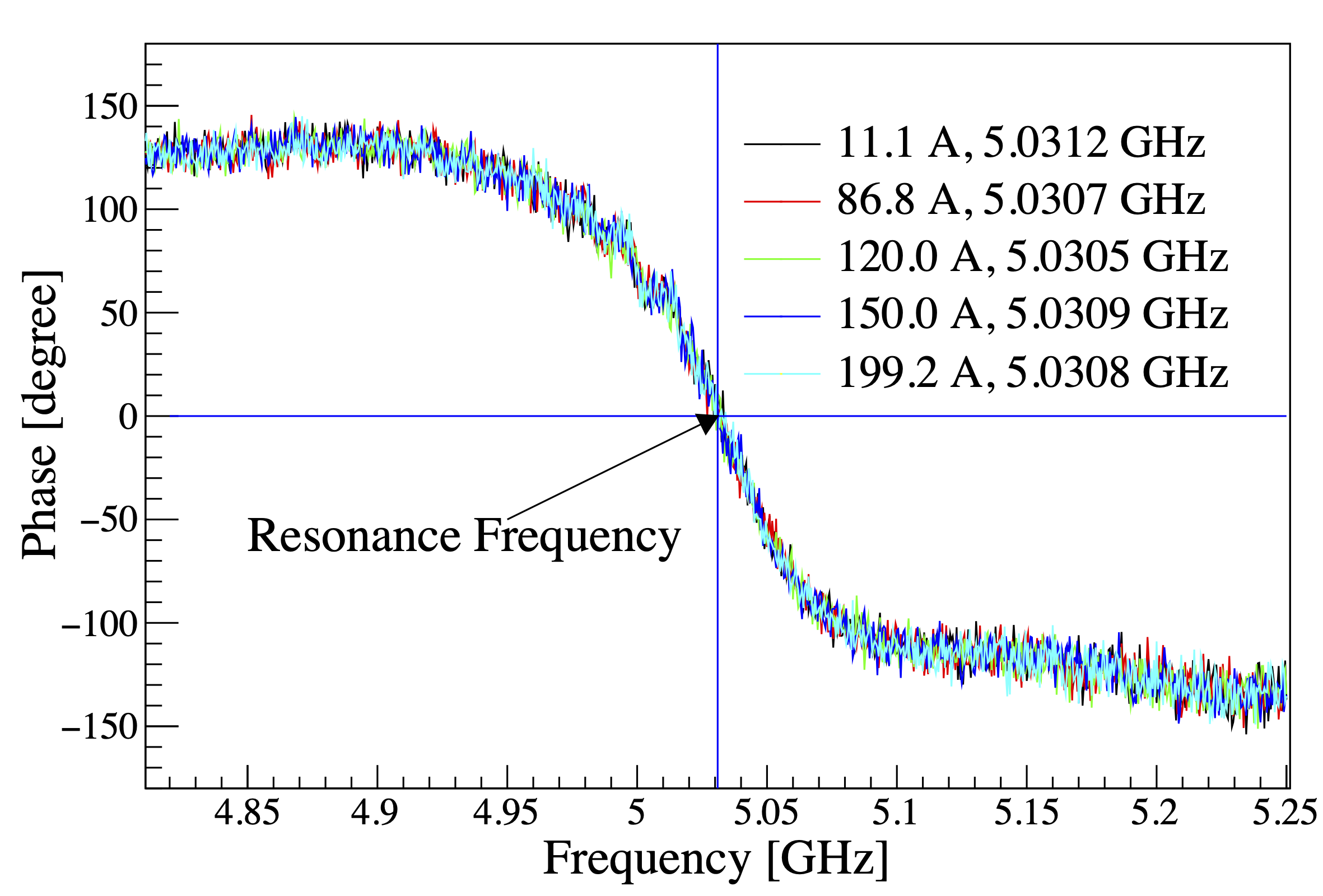}
\caption{JPC phase responses during the 18\,T magnet ramping and dynamic field cancellation by the FCC. The JPC phase responses at different magnet currents are plotted in different colors. No indication of the JPC resonance frequency change is observed.}
\label{fig:jpcphase1}
\end{figure}

\par In practical operations, the FCC is charged with a pre-calibrated current value, which is proportional to the applied current of the 18\,T magnet. The currents of both magnets are controlled using a software package, which sets the target currents and ramping rates, as well as the heater power of the FCC. FIG.~\ref{fig:jpcphase1} shows the phase responses of the JPC during the dynamic field cancellation. The JPC operation and resonant frequency remained stable at 5.03\,GHz until the scheduled ramp-down of the magnet system. The JPC performance was not affected by the ramping process of the 18\,T magnet. This result demonstrates that the residual magnetic field on the JPC is maintained at an acceptable level. \\

\begin{figure*}[t!]
\centering
\includegraphics[width=0.65\linewidth]{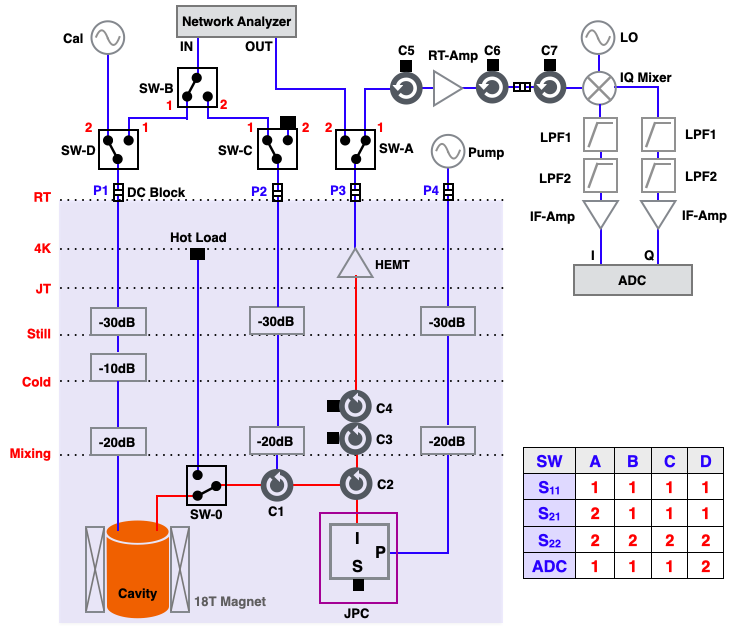}
\caption{Schematic diagram of the microwave circuit. Configurations of the microwave switches for each measurement are listed in the table. Red lines indicate superconducting cables. Blue lines indicate ordinary cryogenic RF cables. Circulators (C1$\sim$C7), amplifiers (JPC, HEMT, RT-Amp, and IF-Amp), low pass filters (LPF1 and LPF2), and the switches (SW-0, A, B, C, and D) are represented by the corresponding icons, respectively. SW-0 is toggled between the cavity and hot-load for $T_S$ calibration. The figure is reprinted from Ref.\,\cite{Lee:2022mnc}.}
\label{fig:capp18trf}
\end{figure*}

\subsection{Receiver Chain and Microwave Circuit}
\par The microwave circuit diagram of the experiment is shown in FIG.~\ref{fig:capp18trf}. There are four RF Input and Output (I/O) channels; weak (P1), coupler (P2), strong (P3), and pump (P4). To prevent DC biases, DC blocks are installed at all four CT-to-RT I/O ports. RT-RF components are connected with 0.086$^{\prime\prime}$ center diameter SMA cables. Axion signals from CT-RF are further amplified by an RT-RF amplifier (Miteq AMF-4F-02000600-13-10P) and an intermediate frequency (IF) amplifier (Mini-Circuits ZX60-100VH+). An IQ mixer (Marki IQ-0307LXP) down-converts the RF signals to the IF frequency by superimposing the RF signal with a Local Oscillator (LO). The mixer splits the signal; one is I (in-phase), and the other is Q (quadrature-phase, a 90$^\circ$ phase shift). These signals from I and Q together keep the phase of the input RF signal. Low-pass filters (LPF) limit the IF bandwidth below 5\,MHz. The filtered I and Q signals are fed to an analog-to-digital-converter (ADC) board for further processing. \\

\par KEYCOM ULT-05 RF cables are used from RT to 4K stages due to their low heat conductivity and insertion loss. The outer surface of these RT to 4K cables is thermally coupled to the 4K stage using copper wires. To suppress any input RT noise, cryogenic attenuators are mounted on each DR stage. KEYCOM ULT-04 cables are used for the weak, coupler, and pump lines below the mixing chamber stage because they are non-magnetic and have high heat conductivity. QUINSTAR QCY-060400CM00 are used as cryogenic circulators (C1, C2, C3, and C4) due to their non-magnetic property. Circulator C1 enables the measurement of the cavity reflection response, and circulator C2 separates the input and output signals of the JPC. Two circulators (C3 and C4) are used as signal isolators required to suppress the reflection signal from the HEMT (Low Noise Factory LNF-LNC2-6A) to the JPC. The hot-load is thermally anchored on the 4K flange, where COAX SC-086/50-CN-CN cables are used between 4K to SW-0 due to their low heat conductivity. NbTi/NbTi SC coaxial cables are used in the CT receiver chain to minimize transmission loss. Two models of SC cables are used, where one is COAX SC-119/50-NbTi-NbTi and the other is KEYCOM NbTiNbTi047. The critical field of NbTi is about 10\,T at 4\,K, so the SC cables remain superconducting under the stray field of the 18\,T magnet (see FIG.~\ref{fig:18TfieldmapXZ}). The transmission loss of the SC cables is less than 0.1\,dB. The combined loss from the cavity to the JPC input is -0.90$\pm$0.02\,dB at 4.8\,GHz, which is incorporated into the calculation of the expected signal power from the cavity. The loss from the JPC output to the HEMT input is -1.00$\pm$0.02\,dB at 4.8\,GHz. The transmission losses of the RF components and cables are measured in a separate CT test setup. The measured transmission loss values are used in the analysis.\\

\begin{figure}[t!]
\includegraphics[width=1\linewidth]{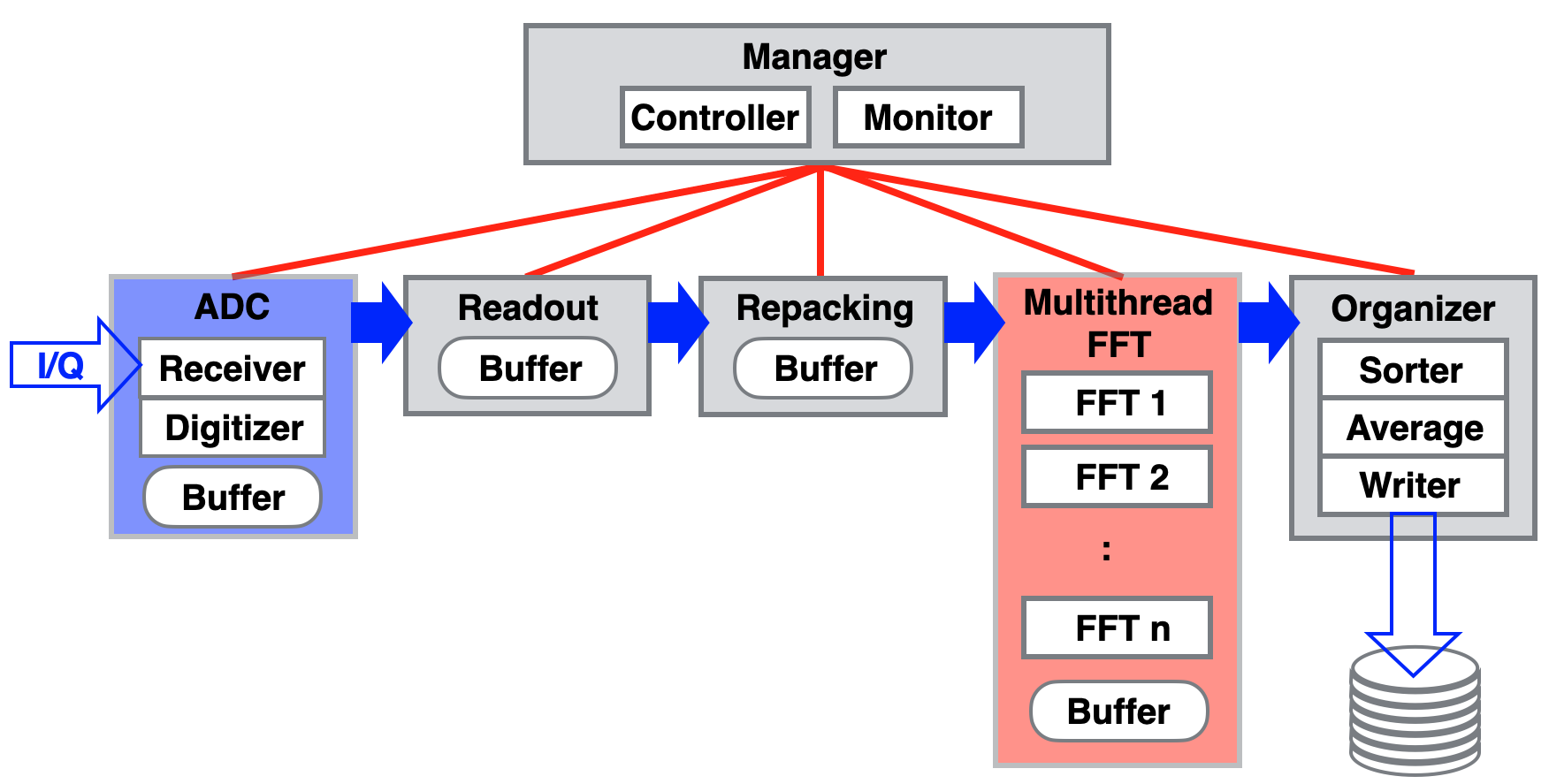}
\caption{Schematic diagram of the DAQ system. Time-domain I/Q signals are digitized in ADC. Readout-Unit reads digitized signals. These signals are buffered in memory for further processing. Repacking process sample data to a proper length for FFT processing. The organizer collects and averages out FFTed power spectra, and stores them on a disk. The manager controls all these processes.}
\label{fig:daq}
\end{figure}

\par  Three signal generators are used to supply RF source: a Keysight N5173B is used for the LO, a Rhode \& Schwarz SMB100A is used to inject calibration signals through the weak port, and a Keysight E8257D is used to provide the pump tone to the JPC. A vector network analyzer (VNA, Rhode \& Schwarz ZNB8) is used to calibrate the RF chain by measuring $\nu_c$, $Q_L$, $\beta$, JPC gain, etc. \\

\par The table in FIG.~\ref{fig:capp18trf} shows the switching combinations for various measurements. The switches are set to $S_{11}$ to measure the reflection response of the weak antenna chain. The switches are set to $S_{21}$ to measure $Q_L$. Signals injected into the cavity through the weak port are read out through the strong port. The switches are set to $S_{22}$ to measure $\beta$. The switching combinations are set to `ADC' for the ADC data-taking mode. SW-0 enables the toggling between the cavity and the hot-load. Calibration signals are injected through the weak port at +250\,kHz detuned from $\nu_C$. All signal generators, VNA, and ADC board are synchronized with a 10\,MHz reference clock, a Stanford Research Systems FS725 rubidium source. Switching combinations for each mode can be controlled by master control software. \\

\subsection{Data Acquisition System}
\par The frontend DAQ consists of an ADC board and a DAQ computer. I and Q signals from RT-RF/IF circuit are fed to the ADC board mounted on the PCIe slot of the DAQ computer. A LabVIEW-based software converts time-domain I and Q data stream to frequency-domain power spectra using a Fast-Fourier-Transform (FFT) algorithm. A schematic diagram of the DAQ is shown in FIG.~\ref{fig:daq}. \\

\par The detector front-end ADC board is a Signatec PX14400A (2-channel, 14-bit, and 400\,MS/s maximum sampling rate). The input voltage range of the ADC can be changed from 220\,mV to 3.5\,V peak-to-peak. The sampling of the ADC is 2-byte (16-bits) and the lowest two bits are always set to 0. The input voltages are digitized to integer numbers from 0 to 65532. The sampling rate can be varied from 20\,MS/s to 400\,MS/s. The ADC has a 512\,MB on-board RAM (Random Access Memory), which is used as a FIFO (First-In-First-Out) buffer in continuous data acquisition mode. The readout software reads the digitized data from the ADC onboard RAM to the memory buffer in the DAQ computer. The DAQ computer is a multi-threaded CPU (two Intel Xeon Gold 6132 with 2.6 GHz, 28 cores, 56 threads). The buffer size is large enough to keep data flow in memory for several minutes, avoiding overflow in the onboard RAM. The repacking software accesses the Readout-buffer to produce 100\,ms of time-bin-slice data for each I/Q channel and stores them in the Repacking-buffer. \\

\begin{figure}[t!]
\includegraphics[width=1.0\linewidth]{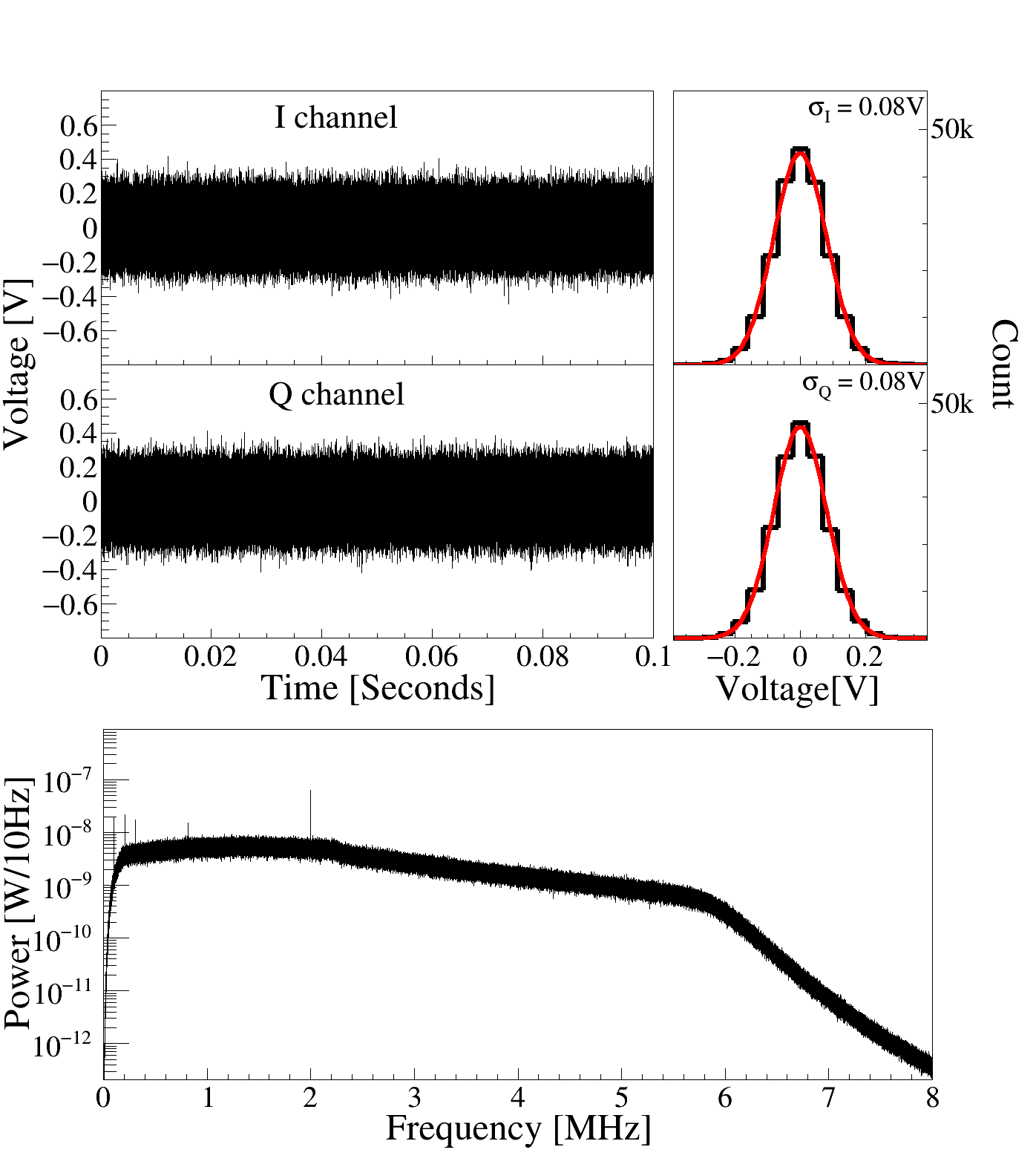}
\caption{Example of time-domain data and FFT-ed frequency-domain power spectrum under low noise conditions. Top: I and Q channel time-domain traces and the voltage distribution of a section of time-domain data. Gaussian fits are overlayed (red curves). The deviations of I and Q voltage distributions under the low-noise condition are about 0.08\,Vs. ADC count is scaled to the corresponding voltage. Bottom: Frequency domain power spectrum in IF frequency after the FFT and image rejection process. Note that the Local Oscillator (LO) frequency is $\nu_C$+2.25\,MHz. The large peak near 2\,MHz is the calibration signal. Several spikes in the spectrum below 1\,MHz are associated with the ground loops and room-temperature electronics. Kink feature in the high-frequency region is due to the 5\,MHz low pass filter.}
\label{fig:IFdata} 
\end{figure}

\par FIG.~\ref{fig:IFdata} shows an example of time-domain I/Q traces and voltage distributions together with an FFT-ed power spectrum. The FFT processes convert the time domain data to the frequency domain power spectra. These FFT processes dominate CPU consumption. 56 threads of parallel FFT processes are used to achieve a dead-time-free DAQ operation. The 100\,ms of time-bin-data corresponds to the 10\,Hz resolution, which is good enough to resolve about a few kHz bandwidths of an axion signal. These FFTed power spectra are stored in the FFT-buffer memory. The I and Q outputs enable image rejection in the IF spectrum\,\cite{AlKenany:2016trt}. A power spectrum of $|S(\omega)|^2$ is defined using the real and imaginary components of the FFTed signals, where $S(\omega)=\{{Re}[I(\omega)]-{Im}[Q(\omega)]\}+i\{{Im}[I(\omega)]+{Re}[Q(\omega)]\}$. The Signal-to-noise-ratio (SNR) of the resulting power spectrum is tested using the source signals. \\

\par The {\it Organizer} reads the FFT-buffer. Every 50 consecutive power spectra (total of 5-sec long time span) are combined as a single averaged power spectrum. The averaged power spectrum is written to the disk, where a typical power spectrum is shown in FIG.~\mbox{\ref{fig:IFdata}} bottom. The overall IF feature is from the low pass filter and the RF components. The first set of time domain data out of every 50 samples is stored for data quality tests and post-analysis. The data rate is 16\,kB/s. These whole processes are controlled and monitored by the {\it Manager} software. The DAQ system is tested as dead-time-free at the sampling rate of below 50\,MS/s. \\

\subsection{System Monitoring}
\par Two main operators are always at the experiment site during the detector operation. However, most of the detector parameters can be remotely monitored and controlled. The DR operation parameters are provided by the DR control software. Eight NI-SCXI-1125 modules collect the control variables; the voltages and shunt current of the 18\,T magnet, the voltages and shunt current of the FCC, the IVC vacuum level, the gate voltage of the HEMT, and the pressure and LHe level of the cryostat. An NI-PXIe-6363 controls the LHe/GHe valves, and a dedicated PC manages the LHe/GHe control loops. A LakeShore-224 measures the temperatures of the cryostat, and a LakeShore-372 resistance bridge measures the temperatures of the cryogenic stages. A Keysight 2182A nanovolt meter is used to read the Hall sensor voltages. These digitized data are transferred to a slow monitor PC, and a LabView software package manages the slow monitor DAQ. \\

\par A total of 104 detector parameters are periodically monitored. The parameters are; 18 DR operation parameters (5 pressures, 5 temperatures of cryogenic stages, 3 heater powers, 1 turbo current, 1 turbo power, 2 turbo temperatures, and 1 turbo speed), 5 temperatures at each cryogenic stage, 1 shunt current of the 18\,T magnet, 1 total voltage and 44 individual DPC voltages of the 18\,T magnet, 2 vacuum level, 1 cryostat pressure, 1 HEMT gate voltage, 1 shunt current and 1 total voltage of the FCC, 2 LHe levels, 10 inner cryostat temperatures, 2 Hall sensors, 1 LHe weight, 2 He gas flows, 2 He mass flows, 4 flow valve status, 1 dewar pressure, 1 DAQ spectrum counter, 1 rotator position, 1 measured cavity frequency, 1 target frequency, and 1 cavity Q-value. A website-based slow monitor displays these parameters in 3-hour and 24-hour time-span graphs. These graphs are updated every 30 seconds. Slow control data with timestamps are stored in a database and transferred to an off-experiment-site computer in real-time.

\section{Experiment\label{Data}}
\par The first CAPP18T dark matter axion search experiment was carried out from the 30th of November to the 24th of December 2020. The scan frequency ranges from 4.7789\,GHz to 4.8094\,GHz (30.5\,MHz bandwidth). There are two data acquisition modes; (1) dark matter axion search and (2) {\it in situ} noise calibration. A master control software automates data acquisition processes in the DAQ computer.    

\subsection{Data Acquisition for Axion Search}
\par For the axion search data acquisition, all detector parameters are fine-tuned and recorded. $\nu_C$ is tuned by the mechanical control of the tuning rod. The coupling of the strong antenna to the cavity is tuned to be critical ($\beta \simeq 1$). The LO frequency is set to $\nu_{\text{LO}}=\nu_C + 2.25$\,MHz. $\nu_J$ is tuned to match with $\nu_C$. The pump frequency and power are tuned to set $G_J$ to be about 27\,dB. The transmission response profile of the cavity, Smith Chart for $\beta$ tuning, and gain profile of the JPC are recorded for data analysis. \\

\par After the detector parameters are set, a 2-min to 15-min long ADC measurement are carried out for the axion dark matter search. The duration of the data taking time is varied depending on the stability of $\nu_C$. During the ADC measurement, a -84\,dBm calibration signal is injected to monitor the stability of the cavity response. The calibration signal frequency is at $\nu_C$ for the first 10\,s, then shifted to $\nu_C$+250\,kHz. \\

\par  After the ADC measurement, a set of cavity parameters ($\nu_C$, $Q_L$ and $\beta$) are measured again to check the stability of the cavity mode. $Q_L$ and $\beta$ are relatively stable compared to $\nu_C$. If $\nu_C$ is off by 10\,kHz from the target frequency, the cavity parameters are tuned, and the ADC measurement is repeated. The data acquired during unstable detector conditions are labeled as {\it unstable}. If $G_J$ at $\nu_C$ is off by 0.5\,dB from the peak gain, $\nu_J$ is retuned to be centered at $\nu_C$. \\

\par In principle, the above data-taking processes are supposed to be repeated until the noise level of the accumulated power spectrum at $\nu_C$ is low enough to probe the target axion signal. In practice, the net exposure time at a given $\nu_C$ is determined based on the detector condition during the operation. The mechanical control of the Kevlar-pulley tuning rod system was vulnerable to vibrations. Especially the LHe transfer to the cryostat introduces a significant long-lasting mechanical turbulence on the system and results in $\nu_C$ drift over 100\,kHz. About 100\,L of LHe is transferred every 22 hours, and it takes about 1\,hour to fill up the cryostat. The instability of $\nu_C$ lasted about 3\,hours even after the LHe transfer was completed. The instability of $\nu_C$ and periodic calibration of detector parameters are the major losses of the axion search livetime. \\

\par Due to the instability of $\nu_C$ at a certain frequency range, two distinct axion search operations were performed: (1) A shallow-scan-sampling in which the total expose time at each target frequency ($\nu_T$) is about 9 minutes. A total of 4.06-days of net shallow-scan data was accumulated. (2) A deep-scan-sampling in which the total expose time at each $\nu_T$ is about 5 to 6 hours. A power spectrum with a significant excess of SNR in the 5\,kHz band over the noise level is qualified as a candidate for the rescanning. The rescanning criteria and analysis of the candidate power spectrum are discussed in the data analysis section.

\subsection{Noise calibration}
\par The system noise temperature is monitored by {\it in situ} measurements during the operation. The total system noise temperature of a receiver chain is given by\,\cite{Friis1944},  
\begin{equation}
T_S = T_P + T_1 + \dfrac{T_2}{G_1} + \dfrac{T_3}{G_1 G_2} + ..., 
\end{equation}
\noindent where $T_P$ is the physical temperature of the system and $T_i$ is the added noise temperature of the $i$-th amplifier with a gain of $G_i$. The physical temperature of the cavity is about 100\,mK. The added noise temperature by the first stage amplifier (JPC) is $T_1\simeq$\,500\,mK and the gain is $G_1\simeq 27$\,dB. The added noise temperature of the second stage amplifier (HEMT) is about $T_2\simeq$\,2\,K, and the gain is $G_2\simeq 37$\,dB. Therefore, the noise contribution from the $T_2$-term is less than 5\,mK, and that from the $T_3$-term is about 0.1\,mK assuming $T_3\simeq$\,300\,K. The RF signal is further amplified by the RT-RF chains. The net RT gain is about 78\,dB (42\,dB of RT-RF gain and 36.5\,dB of IF gain). The expected $T_S$ in an ideal condition is about 625\,mK, where the dominant factors are $T_1$ and $G_1$. In practice, the JPC gain changes depending on the base temperature and frequency. Moreover, the noise power near the cavity resonant frequency with a critically coupled antenna exhibits a frequency-dependent spectral shape. Therefore, $T_S$ is calibrated by {\it in situ} measurements. \\

\begin{figure}[t!]
\centering
\includegraphics[width=0.98\linewidth]{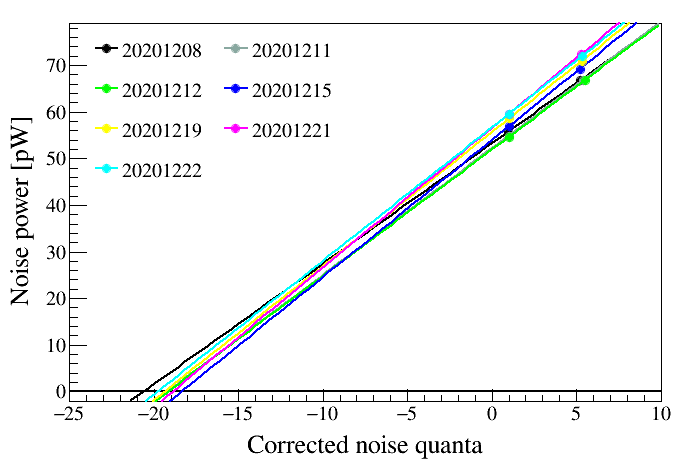}
\caption{Y-factor measurements. Effective temperatures of hot and cold loads are represented in the Johnson noise quanta at the HEMT input port. The intercept of the fitting lines at zero noise power indicates the negative value of the added noise quanta, which is $19.31\pm0.66$.}
\label{fig:Yfactor}
\end{figure}

\par {\it In situ} noise temperature calibration is carried out in two steps using similar methods introduced in Ref.\,\cite{Du:2018uak}; (1) Y-factor measurement to evaluate the noise contribution of the receiver chain without the JPC, and (2) Signal-to-Noise-Ratio-Improvement (SNRI) measurement to obtain the noise power ratio with and without the JPC amplification. \\

\par The Y-factor is defined as the ratio of the noise power of hot ($P_H$) and cold ($P_C$) sources such that $Y = {P_H}/{P_C} = {(T_H + T_A)}/{(T_C + T_A)}$, where $T_{C}$ ($T_H$) is the cold (hot) input noise temperature. For the Y-factor measurement, the JPC pump tone is disabled. The added noise temperature of the RF chain without the JPC is $T_A = {(T_H - YT_C)}/{(Y - 1)}$. The effective input noise temperatures of $T_C$ and $T_H$ on the HEMT are estimated considering the transmission efficiency and thermalization of the RF signal at each DR stage. The physical temperature of the cavity varied from 100 to 120\,mK. $T_C$ at the HEMT input port is estimated from 200\,mK to 240\,mK. The physical temperature of the hot-load varied from 4.4\,K to 5.3\,K, which depends on the operation cycle of the cryostat. $T_H$ at the HEMT input port is estimated to be from 1.2\,K to 1.5\,K. The typical system noise temperature without the JPC is about 4.68\,K. A total of 7 Y-factor measurements were carried out during the operation. FIG.~\ref{fig:Yfactor} shows the results of the Y-factor measurements. $T_A$ is measured to be 4.44$\pm$0.16\,K at 4.8\,GHz (noise quanta of 19.31$\pm$0.66). \\

\begin{figure}[t!]
\centering
\includegraphics[width=0.98\linewidth]{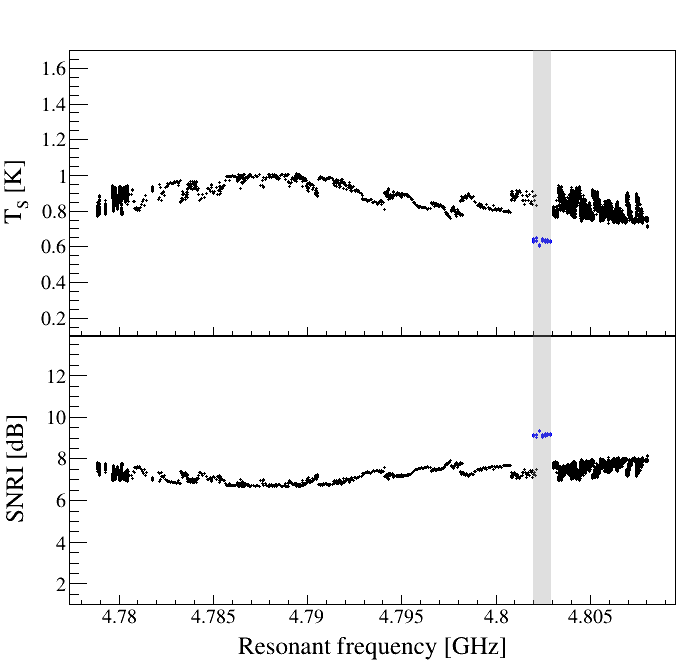}
\caption{Measured $T_S$ and SNRI at $\nu_C$. The frequencies in the vertical gray band were scanned again in August 2021 with an improved thermal contact between the cavity and the tuning rod. The corresponding $T_S$ and SNRI values are shown in blue dots.}
\label{fig:snri}
\end{figure}

\par For SNRI, the noise powers and gain factors of the receiver chain with and without the JPC pump tone are measured. The noise power with JPC-off is $P_{\text{off}} \propto G_{\text{off}} T_R$, where $T_R=T_A+T_C$. The noise power with JPC-on is $P_{\text{on}} \propto G_{\text{on}} T_S$, where $G_{\text{on (off)}}$ is the gain factor of the RF chain in the JPC-on (JPC-off) configuration. The total system noise temperature is
\begin{equation}
T_S = \left[\frac{P_{\text{on}}/G_{\text{on}}}{P_{\text{off}}/G_{\text{off}}}\right] T_R = \frac{T_R}{\text{SNRI}},
\label{eqn:Ts}
\end{equation}
\noindent where SNRI$\,\equiv(G_{\text{on}}P_{\text{off}})/(G_{\text{off}}P_{\text{on}})$. The JPC-on data are obtained for 100 seconds for each data sample. The JPC-off data are obtained 7 times during the operation, and the average value is used. The SNRI values are measured as a function of the frequency. FIG.~\ref{fig:snri} shows $T_S$ and SNRI values at $\nu_C$ during the axion search operation. The SNRI in the frequency region of 4.801\,960 to 4.802\,945\,GHz showed abnormal behavior (gray band), which was caused by an abrupt temperature change in the system due to unexpected cold airflow into the laboratory. The large temperature fluctuations in RT have impacted the DR system and the JPC. Consequently, the resonance frequency shifted down with significant systematics. The corresponding frequency region was scanned again in August 2021 with improved thermal contact between the cavity and the tuning rod. \\

\par
\begin{figure}[t!]
\centering
\includegraphics[width=0.98\linewidth]{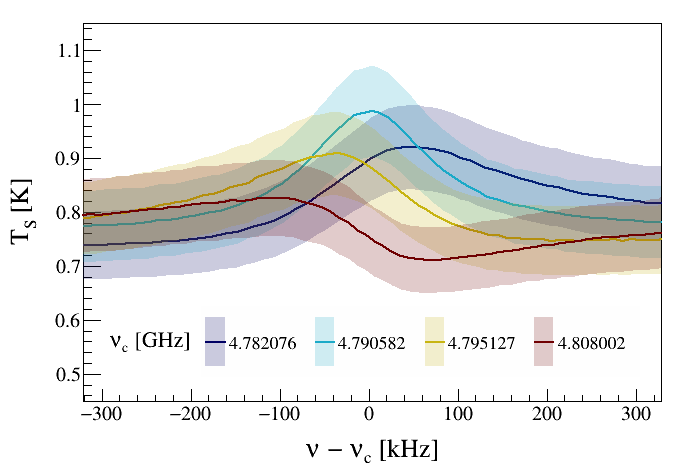}
\caption{Examples of the measured system noise temperatures centered at $\nu=\nu_C$. Lorentzian peaks near $\nu=\nu_C$ are rather anomalous. The origin of the peak is not clearly understood, but it is suspected that they arose from the thermal photons of the hot-tuning rod in the cavity or impedance mismatching between the cavity and the RF components. The amplitude of the hot-peak is reduced in the later measurements (higher frequencies) and becomes asymmetric.}
\label{fig:Tsys}
\end{figure}

 FIG.~\ref{fig:Tsys} shows examples of the measured $T_S$ as a function of $\nu-\nu_C$. The Lorentzian feature at the resonant frequency in FIG.~\ref{fig:Tsys} is unexpected. A similar observation has been reported in Ref.\,\cite{Brubaker:2017rna}, where the feature is assumed to be caused by an imperfect thermal anchoring of the tuning rod with the cavity. In our experiment, the feature gradually changed during the operation and showed asymmetric RF responses. This frequency dependence of the noise temperature is not fully understood. The measured noise temperature at the off-resonance is about 750\,mK on average. The noise temperatures at $\nu_C$ are used in the axion search analysis. 

\subsection{Axion Signal Power and Scan Speed}
\par The expected signal power of the dark matter axions at a given $g_{a\gamma\gamma}$ is estimated using Eqn.\,\ref{eq:APower}. The measured magnetic field ($B(r,z)\bold{\hat{z}}$) is used to calculate the volume averaged magnetic field $\langle \bold{B}^2 \rangle V = \sum_i B_i^2 V_i$, where $V_i$ is the $i$-th segment of the cavity volume. $\langle \bold{B}^2 \rangle V$ is calculated at each tuning rod position. The typical $\langle \bold{B}^2 \rangle$ is about (15.66\,T)$^2$. $C_{010}$ is obtained from the simulation study which is about 0.56 $\sim$ 0.58 in the frequency region of interest. $Q_L$ at CT is measured to be about 24\,400 on average with a 12.3\% standard deviation. $\beta$ is about 1.02 on average with a 8.8\% standard deviation. The expected signal power of the KSVZ axions is about 2$\times10^{-23}$\,W at the frequency of 4.8\,GHz. \\

\par The performance of an axion haloscope is often represented by its frequency scan speed. In an RF chain, the SNR is defined as ${\cal R}=P^a/\delta P_{\text N}$, where $\delta P_{\text N}$ is the uncertainty of the noise power. From the Dicke's radiometer equation, the SNR is given as ${\cal R}={P^a}/({k_B T_S})\sqrt{{\Delta t}/{(\Delta \nu_a)}}$\,\cite{Dicke1946a}. The average scan speed is then represented as 
\begin{equation}
\label{eq:ScanSpeed}
\begin{aligned}
\frac{d\nu}{dt} = y_c\, \frac{g^4_{a\gamma\gamma}}{{\cal R}^2}\Big[\frac{\rho_a}{m^2_a} \frac{\omega_0\langle \bold{B}^2 \rangle V}{k_B T_S} \frac{\beta}{1+\beta}C_{nlm}\Big]^2 Q_L Q_a, 
\end{aligned}
\end{equation}
\noindent where $Q_a$ is the axion quality factor, and the coefficient $y_c(=0.79)$ is determined by the frequency tuning step and the analysis-band-width\,\cite{brubaker2018Thesis}.  Using a typical noise temperature of $T_S$=830\,mK at $\omega_0 = 2\pi\cdot4.8$\,GHz, the scan speed of the experiment for the KSVZ axion with a significance of ${\cal R}=5$ is estimated to be about 60\,MHz/year. This remarkable sensitivity above 10\,$\mu$eV of the axion mass range is achieved by adopting the strongest magnetic field compared to any other haloscope detectors. In this first operation, the critical coupling ($\beta\simeq 1$) setup is used. However, from Eqn.\,\ref{eq:ScanSpeed}, the scan speed is maximized at $\beta = 2$, where the scan speed improves by a factor of 1.7 compared to that of the critical coupling. 

\subsection{Power Spectrum}
\par The data analysis of the dark matter axion search aims to find the signature of the Galactic axions in the measured RF power spectrum. The feature of the axion signal would be an excess power of a few kHz bandwidths at the corresponding axion frequency. For a decisive discovery of the dark matter axions, the signal power should be significant enough over the noise level. True axion signals should be persistent over repeated measurements and consistency tests. \\

\par The discovery or exclusion of an axion signal can hardly be claimed if the detector condition is abnormal. Therefore, the final data samples are carefully selected, and only the samples with good detector conditions are considered for the axion search data analysis. In the present detector operation, the calendar time duration of the axion search is 24.5\,days. Among them, the following raw data samples are removed before any further data processing: (1) data tagged as {\it bad} by the onsite operators due to technical issues, (2) data with incomplete file information, and (3) data acquired during LHe transfers. These pre-selection leaves 1\,116\,404.4\,s (12.9\,days) of data. Abnormal SNRIs in the frequency region of 4.801\,960 to 4.802\,945\,GHz are removed, which leaves 1\,063\,957.5s (12.3days) of data. The vibrations of the detector system cause instability of $\nu_C$. Therefore, data belonging to the criteria of $|\nu_C(t_{i+1})-\nu_C(t_{i})| \ge 10$\,kHz are removed, where $\nu_C(t_{i})$ is the resonant frequency measured at $t_i$. This selection leaves 1\,060\,688.8s (12.3\,days) of data. Similarly, any drastic change of $Q_L$ induces large systematic uncertainty in the signal power. Therefore, data samples belonging to $|Q_L(t_i)-Q_L(t_{i+1})|\ge$\,1\,000 are removed. This selection leaves 1\,047\,782.9s (12.1\,days) of data. These final samples are further processed to search for a potential axion signal. Table~\ref{tab:selection} summarizes the data selection efficiencies. The net efficiency of the axion search livetime over calendar time duration in this operation is 49.4\%. The final sample is 1\,065\,290.9s (12.3 days) of data, including the 17\,508.0s (0.2 days) of net data acquired in early August 2021. Note that the most significant inefficiency of the detector livetime is caused by the LHe transfer and the associated instability of the detector ($\sim$30\% of loss).

\begin{figure}[t!]
    \includegraphics[width=0.98\linewidth]{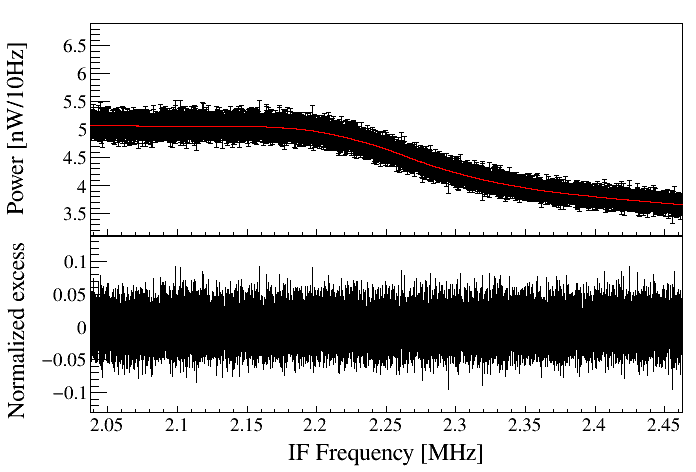}
    \caption{Top: The power spectrum of 3-minute long in IF frequency. The structure of the power spectrum is from the characteristics of the RF components. The red curve is the baseline obtained by the SG-filter. Bottom: Normalized power spectrum. The power spectrum is normalized by the SG-filter baseline, where the mean of the spectrum is shifted to zero.}
    \label{fig:rawspc}
\end{figure}

\begin{table}[b!]
\caption{\label{tab:selection} Data selection criteria and selection efficiency.}
\begin{ruledtabular}
\begin{tabular}{l r r c}
Selection criteria  & Exposure [s] & ([days])  & Efficiency \\
\hline
After pre-selection & 1\,116\,404.4 & (12.9) &  1.000   \\
After anomalous SNRI cut     & 1\,063\,957.5 & (12.3) &  0.953   \\
After $\nu_C$ drift cut      & 1\,060\,688.8 & (12.3) &  0.997   \\
After $Q_L$ fluctuation cut  & 1\,047\,782.9 & (12.1) &  0.988   \\
\hline
2020 net sample  & 1\,047\,782.9 & (12.1) &  0.939   \\
2021 net sample +   &  17\,508.0 & (0.2) &   \\
\hline
Final sample   & 1\,065\,290.9  & (12.3) &   \\
\end{tabular}
\end{ruledtabular}
\end{table}

\par A typical example of the raw power spectrum in IF frequency is shown in FIG.~\ref{fig:rawspc}. The non-uniform spectral shape represents the characteristics of the RF chain response. A Savitzky-Golay filter (SG-filter)\,\cite{SGfilter1964} is applied to obtain the baseline of the power spectrum. The SG-filter is often used for smoothing data to improve analysis precision. The filtering process fits a polynomial function of degree $d$ to 2$W$+1 data points centered at $\nu_s$. After the fit is completed, $\nu_s$ is moved to the center of the next set of 2$W$+1 data points. The process is repeated to the end of the frequency band. In this analysis, $d=3$ and $W=2\,500$ is used. The outputs of the SG-filter are a set of the least-squared polynomial functions that fit the data at the corresponding frequency band. The red curve in FIG.~\ref{fig:rawspc} shows the result of the SG-filter. The raw power spectrum is divided by the SG-filter function to obtain a dimensionless normalized spectrum. The mean of the normalized spectrum is then shifted to zero. The resulting power spectrum is near white noise, and the projected noise spectrum shows a normal distribution. An example of the normalized power spectrum is shown at the bottom panel in FIG.~\ref{fig:rawspc}, with a standard deviation of 0.023. The attenuation of the signal power by the SG-filter is estimated using a simulation study. In the simulation, 10\,000 gaussian random noise samples are generated, and the expected axion signals are mixed in the sample. The significance of the signal power with and without the SG-filter is compared. The attenuation of the axion signal power to noise ratio by the SG-filter with $d=3$ and $W=2\,500$ is estimated to be $\varepsilon_{_\text{SG}}=0.888$. \\

\begin{figure}[t!]
    \includegraphics[width=1\linewidth]{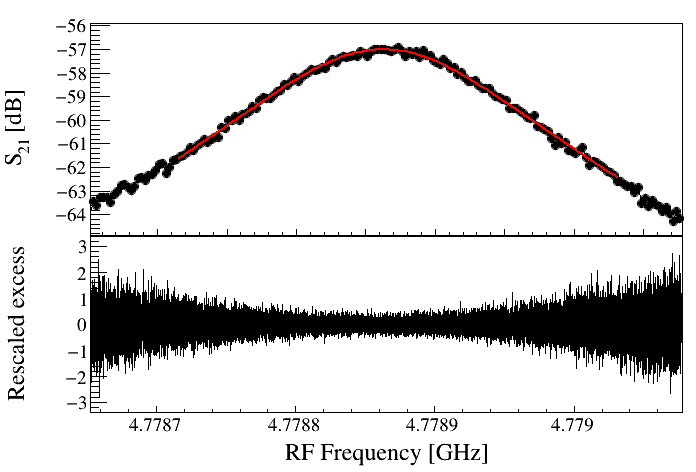}
    \caption{Top: A typical example of $S_{21}$ spectrum. The red curve is a Lorentzian fit. Bottom: Rescaled power spectrum, where the Lorentzian response of the cavity, $T_S$, and input power attenuation ($\xi$) are corrected.}
    \label{fig:s21}
\end{figure}

\par The normalized power spectrum in IF frequency is shifted to its original RF frequency centered at the corresponding resonant frequency. FIG.~\ref{fig:s21} shows an example of an $S_{21}$ measurement. A Lorentzian fit is applied to find $\nu_C$ and $Q_L$. The normalized spectrum is rescaled by $T_S$ and divided by the axion conversion power as
\begin{equation}
P_{i}^r (\nu) = \frac{k_B T_{S} (\nu) \Delta\nu_b P_{i}^n (\nu)} {P_{i}^a (\nu)},
\label{eqn:rescale}
\end{equation}
\noindent where $P_i^n(\nu)$ is the normalized power spectrum of the $i$-th data set. The expected power of the hypothetical dark matter axion is expressed as $P_{i}^a = \xi(\nu) P^a / \cal{L}(\nu)$, where $\xi(\nu)$ is the measured signal attenuation (-0.98\,dB) from the strong antenna to the input port of the JPC, and $\cal{L}(\nu)$ is the normalized Lorentzian curve; ${\cal L}(\nu)=1+4{(\nu - \nu_{c})^2}/{(\Delta \nu_{c})^2}$. The standard deviation of the rescaled power spectrum is given by
\begin{equation}
\sigma_{i}^r (\nu) = \frac{k_BT_S(\nu) \Delta\nu_b \sigma_{i}^n}{P_{i}^a(\nu)},
\label{eqn:rescale_sig}
\end{equation}
\noindent where $\sigma_{i}^n$ is the standard deviation from the Gaussian fit on $P_{i}^n (\nu)$ distribution. 
\begin{figure}[t!]
\includegraphics[width=1\linewidth]{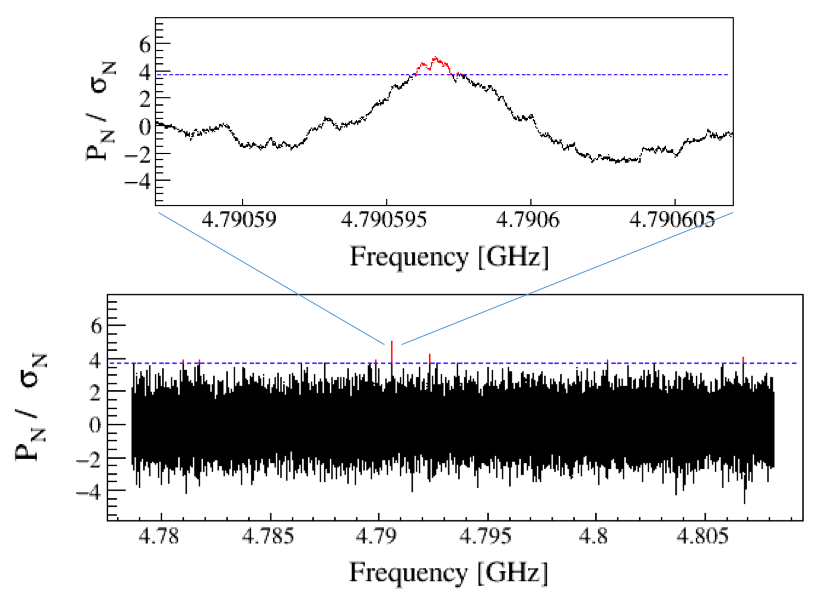}
\caption{Normalized grand power spectrum. The corrected grand spectrum is plotted as a function of frequency and the threshold of 3.718. A total of 8 rescan candidates are found above the threshold. A zoomed-in view around a rescan candidate frequency of 4.790\,597\,GHz is shown at the top plot.}
\label{fig:grandSpect}
\end{figure}
An example of the rescaled power spectrum $P_{i}^r (\nu)$ is shown in the bottom plot of FIG.~\ref{fig:s21}. The shape of the spectrum is largely affected by the $\cal{L}(\nu)$ correction in Eqn.\,\ref{eqn:rescale}. Each bin in the rescaled spectrum forms an independent normal distribution with different standard deviations centered at zero. 5\,906 spectra are combined to form the grand power spectrum. To combine all these spectra, each frequency bin is weighted by the inverse of the corresponding variance. The weighting factor is defined as
\begin{equation}
w_{i}(\nu) = \frac{(\sigma_{i}^r (\nu))^{-2}}{\sum_i (\sigma_{i}^r(\nu))^{-2}}.
\label{eq:weight}
\end{equation}
\noindent The combined grand power spectrum is then obtained by 
\begin{equation}
P_{\text N} (\nu) = \sum_{i} w_{i}(\nu)P_{i}^r (\nu),
\label{eq:sub_wsum}
\end{equation}
\noindent, and the standard deviation of $P_{\text N}$ is given as
\begin{equation}
\sigma_{\text N} (\nu) = \sqrt{\sum_{i} [w_{i}(\nu)\sigma_{i}^r(\nu)]^2}.
\end{equation}
 
\noindent FIG.~\ref{fig:grandSpect} shows the grand power spectrum. The axion dark matter signal search is performed on $P_{\text N}(\nu)$.

\subsection{Systematic Uncertainty}
\par Systematic uncertainties are evaluated on the expected axion signal power. The uncertainty of the magnetic field is less than 1.0\% based on the comparison of the measured field and simulation results. The stability of the magnetic field is better than 0.05\%. The uncertainty of the effective volume of the cavity is negligible. Therefore, the uncertainty of $B^2 V$ is conservatively estimated to be 1.4\%. The uncertainty of $C_{010}$ is about 3.9\%, estimated using the bead perturbation measurement and simulation. $\beta$ is obtained by $S_{22}$ measurements\,\cite{Kajfez1984,ginzton1957microwave}. The input impedance of a resonating system is expressed as ${Z}/{Z_{0}} = j2Q_{1}\delta + {\beta}/{(1+j2(\beta+1)Q_{L}\delta)}$, where $\delta = {(w-w_0)}/{w_0}$ is the frequency detuning parameter, $\omega_0$ is the resonance frequency, and $j2Q_1\delta$ is the linear order term of the Taylor expansion of the external circuit reactance. Plotting the impedance on a Smith chart forms two circles with one circle in another. Data points of the off-resonance frequency form the outer circle with a radius of $R_o$ governed by $j2Q_1\delta$, while data points of near-resonance form the inner circle with a radius of $R_i$ expressed by ${\beta}/{(1+j2(\beta+1)Q_{L}\delta)}$ term. Then $\beta = {(2/D-1)^{-1}}$, where $D = 1 + {(R_i-d_i)}/{R_o}$ and $d_i$ is the distance between the small circle center and origin. The least-square circular fit is applied to the inner and outer circle data to derive $R_i$, $R_o$, and $d_i$. The systematic uncertainty of $\beta$ is estimated by simulations testing the circular fitting method. For an arbitrary $\beta_{T}$, the theoretical input impedance model function of ${Z}/{Z_{0}}$ is plotted on the Smith chart. The systematic deviation of $\beta_T$ and the fit value $\beta_{f}$ is about 0.7\%. The contribution of the $\beta$ uncertainty on the expected axion signal power is $0.4\%$. \\

\begin{table}[b!]
\caption{\label{tab:SysUncert} Systematic uncertainties. The total uncertainty is modeled on the expected axion signal from the cavity.}
\begin{tabular}{l c r}
\hline 
\hline 
~~{Source}                 &{~~~~~~~~~~~~}&{Fractional uncertainty on $P_a$~~}\\
\hline
~~$B^2 V$                  &  &1.4\%~~~~~~~~~~~~~~~~~ \\
~~$Q_L$                    &  &0.5\%~~~~~~~~~~~~~~~~~ \\
~~Coupling ($\beta$)       &  &0.4\%~~~~~~~~~~~~~~~~~ \\
~~Form factor ($C_{010}$)  &  &3.9\%~~~~~~~~~~~~~~~~~ \\
~~$T_S$                    &  &8.5\%~~~~~~~~~~~~~~~~~ \\
\hline 
~~Total                    &  &9.5\%~~~~~~~~~~~~~~~~~ \\  
\hline 
\hline
\end{tabular}
\end{table}

\par The uncertainty of $Q_L$ is correlated with $\nu_C$, and the instability of $Q_L$ and $\nu_C$ are the sources of the uncertainty of the Lorentzian profile. These uncertainties are propagated to the expected axion signal power at the target frequency $\nu_a$. The combined uncertainty of $Q_L$ and $\nu_C$ on the axion signal power is 0.5\%. The systematic uncertainty of $T_S$ is estimated by the Y-factor (3.4\%) and SNRI (7.8\%) measurements. The uncertainty of $T_S$ at $\nu_C$ is 8.5\%. Table~\ref{tab:SysUncert} summarizes the major systematic uncertainties on the expected axion signal power. The total systematic uncertainty is estimated to be 9.5\%.

\subsection{Dark Matter Axion Search Analysis}
\par It is assumed that the dark matter distribution in our Galaxy is a spherical halo, the velocity distribution follows the MB model, and 100\% of the dark matter is formed with axions. The local mass density of the dark matter ($\rho=0.45\text{\,GeV/cm}^3$) and the circular velocity of the laboratory frame with respect to the rest frame of the Galaxy center ($v_c=220\text{\,km/s}$) are the model parameters~\cite{LEWIN199687,Sloan2016a}. The spectrum of the dark matter axion as a function of the signal frequency is given as\,\cite{PhysRevD.42.3572}
\begin{eqnarray}
\Phi_{\text{MB}}(\nu) = &&\frac{2}{\sqrt{\pi}}\left(\sqrt{\frac{3}{2}}\frac{1}{r}\frac{1}{\nu_a\left< \mathbf{v}^2 \right>}\right)\sinh\left(3r\sqrt{\frac{2(\nu-\nu_a)}{\nu_a\left<\mathbf{v}^2\right>}}\right)\nonumber \\
&&\times\exp\left(-\frac{3(\nu-\nu_a)}{\nu_a\left<\mathbf{v}^2\right>}-\frac{3r^2}{2}\right),
\label{eq:mbdist}
\end{eqnarray}
\noindent where $\left<\mathbf{v}^2\right>=\left<v^2\right>/c^2$, $\left<v^2\right> = 3v_c^2/2$, and $r=v_s/\sqrt{\left<v^2\right>}\approx\sqrt{2/3}$. The velocity of the laboratory frame against the galactic halo is the velocity of the Sun ($v_s \approx v_c$), where the relative motion of the Earth is ignored. The expected axion signal power with spectral shape ($\Phi_{\text{MB}}(\nu)$) is compared with the grand power spectrum ($P_{\text N}(\nu)$).

\begin{figure*}[t!]
    \includegraphics[width=1\linewidth]{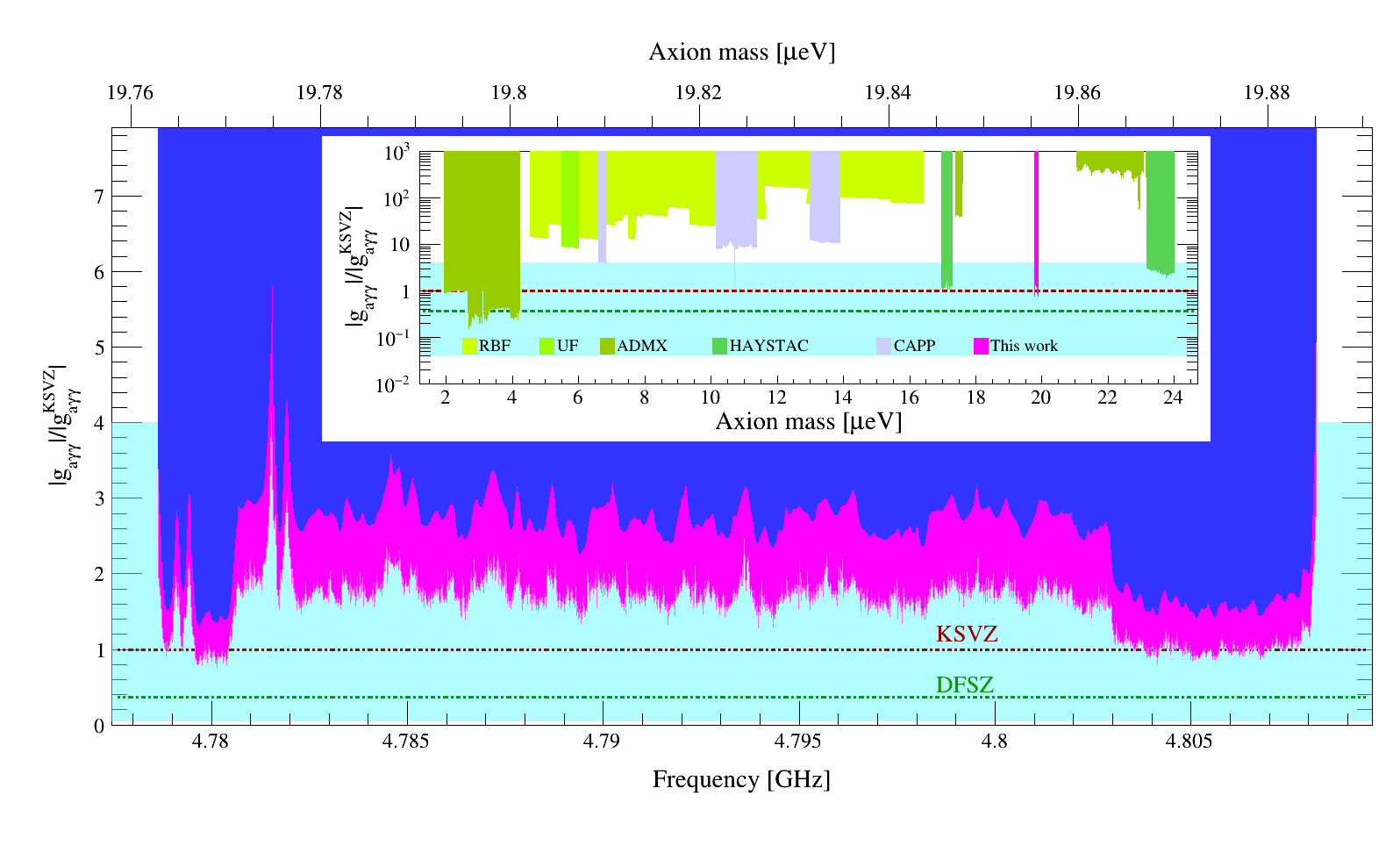}
    \caption{Exclusion limits of the first axion dark matter search by CAPP18T. The Blue shaded region shows the 90\% C.L. exclusion limit using the Frequentist method, which is reported in Ref.\,\cite{Lee:2022mnc}. This analysis follows the HAYSTAC's earlier outline\,\cite{Brubaker:2017rna}. The Magenta shaded region shows the 90\% C.L. exclusion limit using the Bayesian method outlined in Ref.\,\cite{Bartram:2020ysy} by ADMX. The limit of this work (magenta) along with the limits set by the previous haloscope experiments are displayed in the inset figure, with ADMX (olive, Refs.\,\cite{Hagmann1998a,Asztalos:2001tf,Asztalos2002a, Asztalos:2003px, Asztalos:2009yp,Sloan2016a, Boutan2018a,ADMX:2021nhd}, CL\,=\,95\%), HAYSTAC (green, Refs.\,\cite{Brubaker:2016ktl, Backes2021a, PhysRevD.97.092001}, C.L.\,=\,90\%), and CAPP (light purple, Refs.\,\cite{Lee:2020cfj, Jeong:2020cwz, Kwon2021a}, C.L.\,=\,90\%). RBF (lime, Refs.\,\cite{DePanfilis1987a, Wuensch1989a}, C.L.\,=\,95\%) and UF (light green, Ref.\,\cite{Hagmann1990a}, C.L.\,=\,95\%) limits are rescaled based on $\rho_a$\,=\,0.45 GeV/cm$^3$.}\label{fig:limit}
\end{figure*}

\par The statistical fluctuations of the grand power spectrum in the absence of a true signal is a standard normal distribution. An axion signal would appear as an excess of the signal power in a few kHz bandwidths. However, the excess power could also be originated from statistical fluctuation or other unidentified ambient sources. At each frequency band of $k$, a frequency-independent threshold is set by $G_k=\sqrt{R^T/\varepsilon_{_\text{SG}}R^g_k}$, where $R^T$ is the SNR for an axion with frequency-dependent coupling, and $R^g_k$ is the SNR for an axion signal with the KSVZ coupling ($|g^\text{KSVZ}_{a\gamma\gamma}|$). The confidence level at each $k$ is determined where the axion models are ruled out from the function of $|g_{a\gamma\gamma}|$. In this analysis, the SNR threshold is set to $R^T=5.0$, where the corresponding threshold $\Theta$ is $3.718$. The first pass signal scan was carried out to search for the excess signal over the noise level at each data bin $k$. The results identify 8 candidates that exceed $\Theta$. The analysis was repeated using a running-frequency window (RFW) method. In RFW, the search frequency band is fixed as $\Delta \nu = 5$\,kHz for the MB model. The center frequency ran in 10\,Hz steps for each test. The results also identify the same 8 candidate frequencies. \\

\par Likelihood tests were carried out for those 8 candidate power spectra. The expected axion signal shape (Eqn.\,\ref{eq:mbdist}) is compared with each power spectrum within the measurement uncertainty. Simulation studies were carried out to evaluate the probabilities of the axion-like and noise-like signals. The axion-like probability ($p_A$) greater than 1\% is set as the rescan threshold. Table~\ref{tab:candidates} shows the results of the likelihood tests. \\

\begin{table}[b!]
\caption{\label{tab:candidates} Likelihood test results of the 8 excess power spectra. The listed frequency $\nu_a$ is where the signal power exceeds the threshold $\Theta>3.718$. $p_A$ is the axion-like probability and $p_N$ is the noise-like probability. The rescan threshold is set by $p_A$ greater than 1\%.}
\begin{ruledtabular}
\begin{tabular}{c l c c c c}
No. & $\nu_a$ [GHz] & excess [$\sigma$] & $p_A$ & $p_N$ & rescan \\
\hline
1 & 4.780\,998 & 3.881 & 0.0098 & 0.6443 & $\times$\\
2 & 4.781\,756 & 3.870 & 0.0043 & 0.7574 & $\times$\\
3 & 4.789\,855 & 3.928 & 0.0143 & 0.5808 & $\circ$ \\
4 & 4.790\,597 & 5.019 & 0.0682 & 0.2557 & $\circ$ \\
5 & 4.792\,344 & 4.267 & 0.1270 & 0.1519 & $\circ$ \\
6 & 4.793\,604 & 3.754 & 0.1050 & 0.1820 & $\circ$ \\
7 & 4.800\,494 & 3.935 & 0.0098 & 0.6442 & $\times$\\
8 & 4.806\,746 & 4.085 & 0.1569 & 0.1223 & $\circ$ \\
\end{tabular}
\end{ruledtabular}
\end{table}

\par The rescan strategy is based on the fact that a true signal should be persistent at the same frequency ($\nu_a$) across repeated scans. If the candidate signal persists in the rescan data, the signal is classified as the candidate of the dark matter axion signature. An additional test is planned to examine whether the power of the candidate signal would scale as $B^2$, which is expected in Eqn.\,\ref{eqn:pa}. The rescan experiment was carried out in early August 2021. The delayed rescan operation was due to the unexpected relocation of the detector set up at a new experiment site. The five candidate frequencies were carefully investigated. More than factor 2 exposure time was acquired for the rescan data compared to the original data. None of the five candidates shows persistent signals in the rescan data. The absence of a persistent signal after the rescans sets the upper bound on the $|g_{a\gamma\gamma}|$ coupling. \\

\par Assuming a boosted MB distribution of the dark matter axions, the results set the best upper bound of $g_{a\gamma\gamma}$ in the frequency range of 4.7789\,GHz to 4.8094\,GHz ($\Delta f$=30.5\,MHz). Using the Frequentist method outlined in Ref.\,\cite{Brubaker:2017rna}, the limits in the mass ranges of 19.764 to 19.771\,$\mu$eV (19.863 to 19.890\,$\mu$eV) are at 1.5$\times|g_{a\gamma\gamma}^{\text{KSVZ}}|$ (1.7$\times|g_{a\gamma\gamma}^{\text{KSVZ}}|$), and 19.772 to 19.863\,$\mu$eV at 2.7 $\times|g_{a\gamma\gamma}^{\text{KSVZ}}|$ at 90\,\% confidence level (C.L.), respectively\,\cite{Lee:2022mnc}. Using the Bayesian method outlined in Ref.\,\cite{Bartram:2020ysy}, the limits in the mass range of 19.764 to 19.771\,$\mu$eV (19.863 to 19.890\,$\mu$eV) are at 0.98$\times|g_{a\gamma\gamma}^{\text{KSVZ}}|$ (1.11$\times|g_{a\gamma\gamma}^{\text{KSVZ}}|$), and 19.772 to 19.863\,$\mu$eV is at 1.76 $\times|g_{a\gamma\gamma}^{\text{KSVZ}}|$ with 90\,\% C.L., respectively. FIG.~\ref{fig:limit} shows the results of the first CAPP18T axion search. \\

\par In the current RF setup, the added noise of the 1st stage amplifier (JPC) is the dominant source of the total system noise. Therefore, the axion search sensitivity can be substantially improved by suppressing the added noise of the 1st stage amplifier. An excellent example of ultra-low-noise amplification is the squeezed-state receiver (SSR), surpassing the quantum noise limit. Recently, the application of the SSR to the haloscope has been successfully demonstrated by HAYSTAC collaboration\,\cite{Backes2021a}. A similar advanced noise-free amplifier can be adapted in the CAPP18T RF chain to achieve a plausible scan speed down to the DFSZ axion model.

\section{Summary}
\par We report on the first result of searching for invisible axion dark matter using the CAPP18T haloscope. The detector is built based on a new technology of an 18\,T HTS magnet, which is the strongest B-field magnet ever used in axion haloscope experiments. A JPC is used as the first stage RF amplifier, demonstrating a near quantum-limited low-noise performance. A dynamic cancellation of the stray B-field from the 18\,T magnet at the location of the JPC is achieved using an FCC. The JPC is successfully operated under this condition with a typically added noise temperature of about 500\,mK and a gain of 27\,dB. The total system noise temperature is about 830\,mK on average. The axion search data were collected from the 30th of November to the 24th of December 2020, and in early August 2021 for rescan and additional sampling. The net integration time of the axion dark matter search data is 12.3 days. No significant signal consistent with the Galactic dark matter axion is observed. Using a conventional Bayesian method, the results set the best upper bound of the axion-photon-photon coupling in the mass ranges of 19.772 to 19.863\,$\mu$eV at 1.76 $\times|g_{a\gamma\gamma}^{\text{KSVZ}}|$ and 19.764 to 19.771\,$\mu$eV (19.863 to 19.890\,$\mu$eV) at 0.98$\times|g_{a\gamma\gamma}^{\text{KSVZ}}|$ (1.11$\times|g_{a\gamma\gamma}^{\text{KSVZ}}|$) at 90\% CL, respectively. The demonstrated detector performance of CAPP18T will accelerate the worldwide efforts to search for dark matter axions.

\section{Acknowledgement}
\indent This research is supported by the Institute for Basic Science (IBS-R017-D1-2021-a00/IBS-R017-G1-2021-a00). This work is also supported by the New Faculty Startup Fund from Seoul National University. We thank the team of the Superconducting Radio Frequency Testing Facility at Rare Isotope Science Project at IBS for sharing the Helium plant and the experiment space. We also thank Prof.\,Jun-Sung Kim and staff at AMES of POSTECH for calibrating Hall sensors in a low current mode. 

\bibliography{capp18tdetector}
 
\end{document}